\documentclass[11pt,a4paper]{article}
\pdfoutput=1
\usepackage{jcappub}
\usepackage{bm}
\newcommand{\Mpl}{M_{\rm pl}}

\newcommand{\A}{\mathcal A}
\newcommand{\B}{\mathcal B}
\newcommand{\C}{\mathcal C}

\newcommand{\ep}{\epsilon}
\newcommand{\epp}{\epsilon_\phi}
\newcommand{\epc}{\epsilon_\chi}
\newcommand{\etp}{\eta_\phi}
\newcommand{\etc}{\eta_\chi}
\newcommand{\xip}{\xi_\phi}
\newcommand{\xic}{\xi_\chi}
\newcommand{\fnl}{f_{\mathrm{NL}}}
\newcommand{\tnl}{\tau_{\mathrm{NL}}}
\newcommand{\gnl}{g_{\mathrm{NL}}}
\newcommand{\be}{\begin{equation}}
\newcommand{\ee}{\end{equation}}
\newcommand{\bea}{\begin{eqnarray}}
\newcommand{\eea}{\end{eqnarray}}
\newcommand{\eref}{\eqref}
\newcommand{\ds}{\displaystyle}

\graphicspath{{./}}
\notoc

\begin{document}

\title{Large trispectrum in two-field slow-roll inflation}

\author[a]{Joseph Elliston,}
\author[b]{Laila Alabidi,}
\author[a]{Ian Huston,}
\author[a]{David Mulryne}
\author[a]{and Reza Tavakol}

\affiliation[a]{Astronomy Unit, Queen Mary University of London, Mile End Road, London, UK}
\affiliation[b]{Yukawa Institute for Theoretical Physics, Kyoto University, Kyoto 606-8502, Japan}

\emailAdd{j.elliston@qmul.ac.uk}
\emailAdd{laila@yukawa.kyoto-u.ac.jp} 
\emailAdd{i.huston@qmul.ac.uk}
\emailAdd{d.mulryne@qmul.ac.uk}
\emailAdd{r.tavakol@qmul.ac.uk}

\abstract{
We calculate the conditions required to produce a large local trispectrum
during two-field slow-roll inflation. This is done by extending and simplifying the 
`heat-map' approach developed by Byrnes {\it et al}.
The conditions required to generate a large trispectrum are broadly the same as 
those that can produce a large bispectrum. We derive a simple relation between
$\tnl$ and $\fnl$ for models with separable potentials,
and furthermore show that $\gnl$ and $\tnl$ can be related in specific circumstances. 
Additionally, we interpret the heatmaps dynamically, showing how 
they can be used as qualitative tools to understand the evolution of 
non-Gaussianity during inflation. 
We also show how $\fnl$, $\tnl$ and $\gnl$ are sourced by generic shapes
in the inflationary potential, namely ridges, valleys and inflection points. 
}


\maketitle
\newpage
        
\section{Introduction} \label{sec:introduction}

Observational constraints on the statistics of the primordial curvature
perturbation provide a powerful test of inflationary models.
For models driven by canonical scalar fields, the distribution of 
the curvature perturbation is  
extremely close to Gaussian at horizon crossing \cite{Maldacena:2002vr,Seery:2005gb,Seery:2006vu,Seery:2008ax}. 
Moreover, after horizon crossing, the curvature perturbation is conserved for
single-field models \cite{Bardeen:1983qw, Lyth:1984gv, Rigopoulos:2003ak,Lyth:2004gb,Weinberg:2004kf,Langlois:2005ii,Langlois:2005qp}.
In multi-field models, however, if isocurvature modes 
play a role, it can evolve, and this evolution must be 
followed (for example see \cite{Salopek:1988qh, Polarski:1994rz, Sasaki:1995aw,
GarciaBellido:1995qq, Salopek:1995vw, Langlois:1999dw, Gordon:2000hv, Langlois:2010vx, Mulryne:2009kh, Mulryne:2010rp,Dias:2011xy}).
As a result the statistics may become less Gaussian, and 
even sufficiently non-Gaussian to be detectable by 
future probes, such as the Planck satellite \cite{:2006uk}.
In this paper, we consider models in which the non-Gaussianity of the curvature 
perturbation can evolve to become large
during a slow-roll inflationary phase \cite{Rigopoulos:2005us,Vernizzi:2006ve,Alabidi:2006hg,Byrnes:2008wi}. 
Moreover, we focus on two-field 
separable potentials, for which analytic formulae are available 
for the non-Gaussianity parameters \cite{Lyth:2005fi,Vernizzi:2006ve,Seery:2006js,Byrnes:2006vq,Meyers:2011mm}.

The best constraints on local non-Gaussianity to date are provided by the analysis of data 
from the WMAP satellite. The bispectrum is parametrised by $\fnl$, which is
constrained as $-5 < \fnl < 59$ at $95 \%$ C.L. \cite{Komatsu:2010fb}.
$\tnl$ and $\gnl$ parametrise the trispectrum and are constrained as
$-0.6 < \tnl/10^4 < 3.3$ and $-7.4 < \gnl/10^5 < 8.2$ \cite{Smidt:2010ra, Smidt:2010sv}, 
with Ref. \cite{Fergusson:2010gn} finding the slightly different constraint $-5.4 < \gnl/10^5 < 8.6$.
Planck will improve on these constraints considerably, and in the 
absence of a detection is expected to give the bounds 
$|\fnl| < 5$, $\tnl < 560$ \cite{Komatsu:2010hc} and
$|\gnl| < 1.6 \times 10^5$ \cite{Smidt:2010ra}.
If a given inflationary model predicts a magnitude of the $\fnl$, $\tnl$ or $\gnl$ parameters greater 
than these forecast bounds, then we refer to such a prediction as
`observationally relevant'. This is in contrast to a non-Gaussianity 
that could in principle be measured by an ideal observation, which conservatively 
requires these parameters to have 
magnitudes greater than unity. Following the standard diction in the literature, 
we describe such models as producing a `large' non-Gaussianity. \\

Our aims in this paper are as follows: 
First, to show the constraints on the initial and final conditions of the inflationary 
evolution that lead to large values of the $\gnl$ and $\tnl$ parameters, 
and to compare these to the constraints that are required to produce a large value of $\fnl$.
Secondly, to derive a simple relation between
$\tnl$ and $\fnl$ for two-field slow-roll models with separable potentials, 
which is compatible with the well-known consistency relation \cite{Suyama:2007bg}. 
We will also investigate under which conditions it is possible to 
deduce robust relations between $\gnl$ and $\tnl$.
Thirdly, following earlier studies of the $\fnl$ parameter 
\cite{Elliston:2011dr,Elliston:2011et},
we discuss the behaviour of $\tnl$ and $\gnl$
associated with generic features in inflationary potentials, namely
ridges, valleys and inflection points.
Finally, we demonstrate that, in addition to showing the spectrum of 
possible behaviours, our formalism can be used as a qualitative tool to give
insight into how the non-Gaussianity parameters evolve during inflation.

Our method is an extension of the graphical approach employed 
by Byrnes {\it et al.} \cite{Byrnes:2008wi} to study the $\fnl$ parameter.
The method begins by re-casting the expressions for  
$\fnl$, $\tnl$ and $\gnl$ as a sum of terms. Each term is the product of a combination 
of slow-roll parameters and a function of two other parameters related to the initial 
and final conditions satisfied by a given evolution. Since 
the combination of slow-roll parameters is guaranteed to be much less than unity, 
if the non-Gaussianity parameters are to be large,  
the functions must take values much greater than unity.
One can therefore make 2D contour plots (or `heatmaps') of these functions, 
and so identify the regions in parameter space which can give rise to large local 
non-Gaussianities. In this study, we have found that it is possible to simplify the 
expressions to such an extent that only one heat-map is required to understand 
the conditions for a large bi-spectrum, one additional heat-map is required for $\tnl$, 
and three further heatmaps are required for $\gnl$. 

Recently, Peterson \& Tegmark have also undertaken a study of the bi- and tri-spectrum
parameters \cite{Peterson:2010np,Peterson:2010mv,Peterson:2011yt} in the 
setting of slow-roll inflation, arriving at
compact relationships between $\tnl$, $\gnl$, $\fnl$ and the tilts of the 
curvature and isocurvature power spectra. 
Our approach is complimentary to their study, and provides new insights. 
For example, working directly with analytic expressions for $\fnl$, $\tnl$ and $\gnl$ 
allows us to present more explicit formulae, and in particular helps us to understand 
the evolution of non-Gaussianity directly in terms of the shape of the inflationary 
potential. Our general conclusions are, however, very similar.

The paper is structured as follows: \S \ref{sec:theory} provides some background material
and reviews how analytic expressions for observable parameters can be derived 
using the $\delta N$ formalism. 
The expressions themselves are given in appendix \ref{appendixA:fullcalcs}. \S \ref{sec:analytics} 
shows how we can simplify these expressions for the bi- and tri-spectra which we then analyse
in \S \ref{sec:bispectrum_analysis} and \S \ref{sec:trispectrum_analysis} respectively.
This analysis motivates three generic shapes in the inflationary potential 
which we study in \S \ref{sec:transient_calcs}, finding the peak values of $\fnl$, $\tnl$ and $\gnl$ 
in each case. In \S \ref{sec:models} we give concrete examples which illustrate the 
findings of this paper. 
We conclude in \S \ref{sec:conclusions}.

\section{Background Theory} \label{sec:theory}
We consider inflation driven by two   
canonical scalar fields $\phi_i$ with $i=1,2$,
self-interacting through a potential $W(\phi_1 ,\phi_2$).
The scalar field equations of motion are 
\be
\label{eq:eoms}
\ddot \phi_i + 3H \dot \phi_i + W_{,i} = 0\,,
\ee
where a comma denotes partial differentiation with respect 
to the field $\phi_i$, dots are derivatives with respect to proper 
time $t$, and $H$ is the Hubble rate given
by the Friedman equation
\be
3 \Mpl^2 H^2 = W + \frac{1}{2} \sum_{i} \dot \phi_i^2 \,.
\ee
We choose definitions of the slow-roll parameters as
\be
\begin{array}{r l l l r l l}
\epsilon_i &=& \ds{\frac{\Mpl^2}{2} \frac{W_{,i}^2}{W^2} } \,,& \qquad &
\epsilon &=& \ds{\sum_{i=1}^2 \epsilon_i}\,, \\
\eta_{ij} &=& \ds{{\Mpl^2} \frac{W_{,ij}}{W}} \,, & \qquad &
\xi^2_{ijk} &=&\ds{{\Mpl^3} \sqrt{2 \ep} \frac{W_{,ijk}}{W}}\,.
\end{array}
\label{eq:sr_parameters}
\ee
Expressions for $\xi^2_{ijk}$ and $\eta_{ij}$ in 
the frame adapted to the field evolution, which we refer to as the 
kinematic frame, are given
in appendix \ref{appendixA:fullcalcs}. Inflation occurs when $\epsilon \lesssim 1$, and
the so called 
`slow-roll limit' is given by $\epsilon \ll 1$. In this limit the 
fields' kinetic energy can be
neglected, and the
field equations are well approximated by the slow-roll equations of motion
\be
\label{eq:sr_eoms}
3H \dot \phi_i + W_{,i} = 0\,, \qquad3 \Mpl^2 H^2 = W \,.
\ee

\subsection{The Curvature Perturbation}
The primordial curvature perturbation on uniform density spatial hypersurfaces (see for example \cite{Malik:2008im})
is denoted by $\zeta$.

The statistical properties of $\zeta$, which are constrained by 
observations, are commonly measured in terms of its  
power spectrum, bispectrum, and trispectrum, 
defined as 
\begin{eqnarray}
\label{spectra}
\langle\zeta_{\mathbf k_1}\zeta_{\mathbf k_2}\rangle &\equiv& (2\pi)^3
\delta^3 ({\mathbf k_1} + {\mathbf k_2}) \frac{2\pi^2}{{k_1}^3}{\cal P}_{\zeta}(k_1) \, , \nonumber \\
\langle\zeta_{\mathbf k_1}\,\zeta_{\mathbf k_2}\,
\zeta_{\mathbf k_3}\rangle &\equiv& (2\pi)^3 \delta^3 ( {\mathbf k_1}+{\mathbf k_2}+
{\mathbf k_3}) {\cal B}_\zeta( k_1,k_2,k_3) \,,\nonumber\\
\langle\zeta_{\mathbf k_1}\,\zeta_{\mathbf k_2}\,
\zeta_{\mathbf k_3} \, \zeta_{\mathbf k_4} \rangle &\equiv& (2\pi)^3 \delta^3 ( {\mathbf k_1}+{\mathbf k_2})+
{\mathbf k_3}+{\mathbf k_4}) {\cal T}_\zeta( k_1, k_2, k_3,k_4) \,,
\end{eqnarray}
respectively. For the local shape of non-Gaussianity, 
it is convenient to further parametrise 
the bispectrum and trispectrum in terms of the dimensionless parameters,  
$\fnl$, $\tnl$ and $\gnl$, defined by
\bea
\label{fnl}
{\cal B}_{\zeta}(k_1,k_2,k_3)&=&\frac{6}{5} \fnl \left [P_\zeta(k_1) P_\zeta(k_2) 
+ \mathrm{3~perms} \right]\,, \\
\label{tnlgnl}
{\cal T}_\zeta( k_1, k_2, k_3,k_4) &=& \tnl \left [ P_\zeta(k_{13}) P_\zeta(k_3) P(k_4) + \mathrm{11~perms}\right] \nonumber\\
&& \quad+ \frac{54}{25}\gnl \left [ P_\zeta(k_2)  P_\zeta(k_2) P_\zeta(k_4) +\mathrm{3~perms} \right]\,.
\eea

\subsection{The $\delta N$ formalism}

In order to follow the evolution of $\zeta$ on super-horizon scales, and 
calculate its statistics, we employ the separate universe approach to perturbation 
theory \cite{Lyth:1984gv,Wands:2000dp}, and the $\delta N$ 
formalism \cite{Starobinsky:1986fxa,Sasaki:1995aw,Lyth:2005fi}.  
In this approach, spatial gradients are neglected on scales greater than the horizon size,
and each spatial point is assumed to evolve as a separate FRW universe.
Choosing a flat initial slicing at the time at which observational scales 
crossed the cosmological horizon, $t=t^*$, and a later uniform density 
(constant $H$) slicing at $t=t_c$, then $\zeta$ on the final slicing can be 
equated with the excess expansion, $\zeta = \delta N$.

During slow-roll inflation, field velocities are
functions of field positions. Taking this to be a good
approximation at horizon crossing, the subsequent number of efolds
undergone by any `separate universe' is 
a function purely of the initial field values on the flat slicing, $N(\phi_1^*,\phi_2^*)$,
even if the evolution subsequently evolves
away from slow-roll. One therefore finds that  
\be
 \label{eq:deltaN}
\zeta \equiv \delta N = N_{i} \delta \phi_i^*
+ \frac{1}{2!}  N_{ij} \delta \phi_i^* \delta \phi_j^* + \frac{1}{3!} N_{ijk} \delta \phi_i^* \delta \phi_j^* \delta \phi_k^* + \dots\,,
\ee
where here and throughout we employ the summation convention,  
$N$ is the number of efolds
from $t^*$ to $t_c$, a subscript $i$ on $N$ represents
a derivative with respect to the light fields at horizon crossing 
$\phi^*_i$, and $\delta \phi_i^*$
are the field fluctuations on a flat hypersurface at horizon crossing.

Using the expression for $\zeta$ from Eq.~\eref{eq:deltaN}, and 
recalling Eqs. \eqref{spectra}-\eqref{tnlgnl}, one can write observational 
parameters in terms of the derivatives of $N$ \cite{Sasaki:1995aw,Lyth:2005fi,Seery:2006js,Byrnes:2006vq},
\be
P = N_i N_i { P}_* \,, \qquad n_s -1=\frac{2}{H^*}\frac{\dot{\phi}^*_i N_{ij} N_j}{(N_k N_k)^2}
-2 \epsilon^* \,, 
\ee
\be
\frac{6}{5} \fnl = \frac{ N_i N_j N_{ij}}{\left(N_k N_k \right) ^2} \,, \qquad
\tnl=\frac{ N_{ij} N_{ik} N_{j} N_k }{\left(  N_{k}N_k \right) ^3}\,, 
\qquad \frac{54}{25} \gnl = \frac{ N_{ijk} N_i N_j N_k }{\left(  N_k N_k \right) ^3}\,,
\ee
where $n_s$ is the spectral index, $P_* = H_*^2/(2\pi)^2$, and where we have assumed the field 
fluctuations to be Gaussian at horizon crossing, which we recall is an excellent 
approximation for canonical fields \cite{Seery:2005gb,Seery:2006vu,Seery:2008ax}.

\subsection{Analytic formulae}

To calculate the observational parameters for a given inflation model we require the 
derivatives of $N$. These can always be calculated numerically, but 
analytic progress is possible only when the slow-roll equations of motion,
Eq.~\eref{eq:sr_eoms}, are a good approximation, and when a special `separable'
form of potential is assumed. This can either be a sum-separable potential of the form 
$W = \sum_i V_i(\phi_i)$ \cite{GarciaBellido:1995qq, Vernizzi:2006ve, Battefeld:2006sz, Seery:2006js}, 
or a product-separable form, $W = \Pi_i V_i(\phi_i)$ \cite{Choi:2007su}. 
It is also possible to extend the analytic analysis slightly to models of the form 
$W = (\sum_i V_i(\phi_i))^{1/A}$ \cite{Wang:2010si}, where
$A$ is an arbitrary constant. For simplicity, we will restrict our study to 
two-field sum- and product-separable potentials, labelling the fields 
$\phi_1 = \phi$ and $\phi_2 = \chi$. 
The two-field analytic formulae we employ for the derivatives of $N$, and for
the observation parameters which are derived from them, are given in appendix \ref{appendixA:fullcalcs}.

\subsection{The evolution of statistics and the adiabatic limit} 

During inflation, if a given field space evolution follows a 
straight trajectory, then 
the statistics of the curvature perturbation will remain constant.
Conversely, if the trajectory turns, the statistics will 
evolve \cite{Gordon:2000hv,Rigopoulos:2005us,Vernizzi:2006ve}. 
During a turn, typically in the early stages, it is possible for the statistics of the curvature 
perturbation to become highly non-Gaussian \cite{Alabidi:2006hg,Byrnes:2008wi,Byrnes:2008zy}. 
If inflation were then to end suddenly, perhaps through a hybrid transition, it 
is \emph{possible} that this large non-Gaussianity is preserved into the
subsequent phases of the universe's evolution. Often, however, if the turn is completed, 
the non-Gaussianity returns to negligible levels. This possibility was emphasised 
in Ref.~\cite{Meyers:2010rg, Meyers:2011mm}. 

This is not the only possibility, however. It is also possible for the 
bispectrum \cite{Kim:2010ud, Elliston:2011et, Elliston:2011dr,Mulryne:2011ni} 
and trispectrum \cite{Kim:2011je} to be large once
a turn is completed, even after all isocurvature 
perturbations have decayed. If isocurvature modes decay, $\zeta$ and its 
statistics are subsequently conserved on super-horizon scales \cite{Rigopoulos:2003ak,Lyth:2004gb}, 
and this has recently been referred to as the adiabatic limit \cite{Elliston:2011et,Seery:2012vj}. 
As discussed extensively in the literature (for example see 
\cite{Polarski:1994rz, GarciaBellido:1995qq, Langlois:1999dw, Elliston:2011et}),
reaching such a limit greatly simplifies the task of making observational predictions for 
an inflationary model. We emphasis, however, that there is no 
requirement that this occur, and if it does not, isocurvature 
modes will persist at the  
end of inflation \cite{Huston:2011fr}.
The decay of isocurvature modes during slow-roll inflation occurs if the evolution reaches a 
regime which is effectively single field, such as a valley region with steep sides which 
force isocurvature perturbations to decay. 
One purpose of our study is to systematically identify all of the conditions 
which lead to a large $\tnl$ and $\gnl$ during slow-roll inflation, both in cases 
where isocurvature modes have, or have not, decayed.

If the evolution reaches a straight line trajectory along one of the axes, or the 
isocurvature modes decay,  
there is a considerable simplification of the formulae for $\fnl$, $\gnl$ and $\tnl$  
given in appendix \ref{appendixA:fullcalcs}. In particular, the terms involving
$Z$, $\A$ and $\B^2$ (or $\A_P$ and $\B_P^2$ ) as defined in Eqs. \eqref{eq:Z}, 
\eqref{eq:ss_A} and  \eqref{eq:ss_B} (or eqs. \eqref{eq:ps_A} and \eqref{eq:ps_B}) 
tend to zero\footnote{As discussed at length elsewhere \cite{Elliston:2011dr},  
there is a possibility that, for sum-separable potentials, $Z$ does not go to zero. 
This exception only occurs if one of the fields, $\phi_k$, is completely orthogonal
to the final straight line direction, \emph{and} $V_k (\phi_k)$ tends to a constant as $V_k (\phi_k)'$ 
tends to zero. Such a possibility can be avoided since we are free to reparametrise the
potential and associate the problematic constant with another of the fields.}. 
These terms vanishing is referred to as the Horizon Crossing Approximation (HCA) \cite{Kim:2006te}.
This greatly simplifies the expressions 
\eqref{eq:ss_params1} and  \eqref{eq:ps_params1}, 
allowing us to make much stronger statements about the relations between the observable 
non-Gaussianity parameters. 

A final clarification is in order. Throughout this paper we calculate statistics 
on uniform density hypersurfaces. If, however, inflation ends suddenly, 
a further complication is that the surface on which inflation ends may not be a uniform density one.
If this is not the case, an additional source of $\zeta$ will be produced \cite{Lyth:2005qk}, which 
will alter the statistics we calculate. 
We do not discuss this possibility further here, except briefly in \S \ref{sec:hybrid}. We note 
that in cases where all isocurvature modes decay there is no such possible additional effect.

\section{Analytic non-Gaussianity} 
\label{sec:analytics}

In this section we present the central results of our paper. 
We reformulate the analytic formulae for the non-Gaussianity parameters, which are reviewed 
in appendix \ref{appendixA:fullcalcs}, into a set of terms, each of which is the 
product of a slow-roll suppressed part, and a function of two variables 
related to the initial and final conditions. Only the functions can be large, 
and so by plotting them as two-dimensional heatmaps, 
we can identify the initial and final conditions which lead to 
large non-Gaussianities. A key part of this procedure involves neglecting terms 
in these expressions that are always too small to contribute significantly to a large
non-Gaussianity. We will see in subsequent sections, that these 
heatmaps are also useful for understanding the dynamical evolution of the 
non-Gaussianity parameters. Byrnes {\it et al.} \cite{Byrnes:2008wi} 
used a similar method to study the $\fnl$ parameter. We first improve this 
bispectrum analysis by reducing to one the number of relevant heatmaps, 
and then consider the trispectrum.

\subsection{Variables}

We begin by defining the first important variable, the angle $\theta$ in 
the $\{\phi$, $\chi\}$ phase space in which the inflationary trajectory evolves. 
This can be done in terms of the slow-roll parameters 
 $\epp$ and $\epc$, defined in Eq.~\eqref{eq:sr_parameters}, as
\be
\frac{\epp}{\ep} =\cos^2 \theta \,, \qquad
\frac{\epc}{\ep} = \sin^2 \theta.
\ee
Since we assume that both fields are monotonically decreasing (which follows 
from our use of the slow-roll equations of motion),
$\theta$ is constrained to lie in the range $0\leq \theta \leq \pi/2$.

The $\delta N$ expressions, given in appendix \ref{appendixA:fullcalcs}, also 
involve the quantities $u$ and $v$, defined in Eqs. \eqref{eq:ss_uv} 
and \eqref{eq:ps_uv}, which lie 
in the range zero to one. It proves convenient, therefore, to define a second angle, 
$\alpha$, in terms of these quantities as 
\be
u = \cos^2 \alpha \,, \qquad
v = \sin^2 \alpha.
\label{eq:alpha}
\ee
We note that in the product-separable case, $\alpha = \theta$.
The situation is not so simple in the sum-separable case, as we shall 
discuss in \S \ref{sec:bispectrum_analysis}.

Substituting these definitions into the expressions for the non-Gaussianity parameters, 
Eqs. \eqref{eq:ss_params} and \eqref{eq:ps_params}, 
we can eliminate the variables 
$u,v,\epp$ and $\epc$ in favour of $\alpha,\theta$ and $\ep$.
The observables are then only functions of 
$\alpha,\theta$ and $\theta^*$, as well as $\ep$ and the other slow-roll parameters.
Some of these slow-roll parameters are evaluated at horizon crossing, whilst others 
are evaluated on a later constant energy density  hypersurface, usually labelled `$c$'.
In the following we can drop this label  without ambiguity, since 
within the $\delta N$ formalism all quantities evaluated after horizon 
crossing are calculated on a uniform density hypersurface.
Quantities without a `$*$' label are therefore 
assumed to be calculated on this later-time uniform density hypersurface.

\subsection{Bispectrum}

For the $\fnl$ parameter the procedure outlined above leads to the expressions 
\be
\begin{array}{l l l l}
\ds{\frac{6}{5} \fnl} &=& f_1 \ep^* - f_2 \eta_{ss}^* + f_3
\eta_{\sigma s}^* + 2 f \, \Omega \, (\eta_{ss}-\ep) & ~~~~~~\mbox{-- Sum
separable} \,, \vspace{2mm}\\
\ds{\frac{6}{5} \fnl} &=& - f_2 \eta_{ss}^* + f_3 \eta_{\sigma s}^* +
2 f  \eta_{ss} & ~~~~~~\mbox{-- Product separable} \,,
\label{eq:fnl_full}
\end{array}
\ee
where, similarly to Ref.~\cite{Byrnes:2008wi}, we have defined the functions $f (\alpha,\theta^*)$, $f_i (\alpha,\theta^*)$, $\Lambda$ and $\Omega$ as
\be
\begin{array}{l l l l l l l}
f &=& \ds{\frac{\sin^2 2\alpha}{4 \Lambda^2} (\cos^2 \alpha - \cos^2 \theta^*)^2 \,,} & \qquad & f_2 &=& \ds{\frac{1}{\Lambda^2} \left( \cos^6 \alpha \sin^4 \theta^* + \sin^6 \alpha \cos^4 \theta^* \right)} \,, \vspace{2mm}\\
f_1 &=& \ds{\frac{\sin^2 2\theta^*}{2\Lambda}}  \,,  & \qquad &
f_3 &=& \ds{\frac{\sin 2\theta^*}{2\Lambda^2} \left( \cos^6 \alpha \sin^2 \theta^* - \sin^6 \alpha \cos^2 \theta^* \right)}  \,, \vspace{2mm} \\
\Lambda &=& \ds{\cos^4 \alpha \sin^2 \theta^* + \sin^4 \alpha \cos^2 \theta^* \,,}& \qquad &
\Omega &=& \ds{\frac{W^2}{W_*^2} \frac{\sin^2 2 \theta}{\sin^2 2 \alpha}} \,. 
\end{array}
\ee

Since $0 \leq \Omega \leq 1$
\footnote{This follows by using \eqref{eq:ss_uv} and the 
associated definitions found in appendix \ref{appendixA:fullcalcs} to find
\be
\Omega = \frac{W^2 \epp \epc}{(U^* \ep +V \epp - U \epc)(V^* \ep - V \epp + U \epc)} 
\leq \frac{W^2 \epp \epc}{(U \ep +V \epp - U \epc)(V \ep - V \epp + U \epc)} = 1 \,,
\label{eq:nifty_relation}
\ee
where the second inequality follows using $U \leq U^*$ and  $V \leq V^*$.},
the functions $f_{1 \to 3}$ and $f$ all multiply quantities of 
$\cal{O} (\epsilon)$ or smaller, and so a necessary, though not sufficient, 
condition for $\fnl$ to be large
is that the magnitude of one or more of these functions is large. 
Our analysis, therefore, identifies only the conditions for which it is \emph{possible}
to produce a large $\fnl$ during slow-roll inflation.

Because we are only interested in cases where $\fnl$ can be large, 
the expressions \eqref{eq:fnl_full} may be further simplified by noting that 
$|f_1|$ is bounded by order of unity and so the term $f_1 \epsilon^*$ is negligible.  
Similarly, the term $f_3 \eta_{\sigma s}^*$ is negligible and can be dropped. 
This latter result follows by recalling that we have the freedom to interchange between 
$\eta_{\sigma \sigma}$, $\eta_{\sigma s}$ and $\eta_{ss}$ via the relations
\be
\begin{array}{llll}
\eta_{\sigma s} &=& \frac{1}{2} \tan 2 \theta \, (\eta_{ss} - \eta_{\sigma
\sigma}) &\qquad \mbox{-- Sum separable} \,, \vspace{2mm}\\
\eta_{\sigma s} &=& \frac{1}{2} \tan 2 \theta \, (\eta_{ss} - \eta_{\sigma
\sigma}+2\ep) &\qquad \mbox{-- Product separable} \,.
\label{eq:eta_relationship}
\end{array}
\ee
Considering first the sum-separable case and expanding $f_3 \eta_{\sigma s}^*$ as
\be
 f_3 \eta_{\sigma s}^* = \Big[\sin 2 \theta^* f_3 \Big] \eta_{\sigma s}^* 
+\Big[\frac{1}{2} (1-\sin 2 \theta^*) \tan 2\theta^* f_3  \Big] (\eta_{ss}^*- \eta_{\sigma \sigma}^*),
\label{eq:trick1}
\ee 
our result follows from the fact the the terms in 
square brackets can never become larger than order 
of unity. 
This procedure works because 
$\eta_{\sigma s}^*$ tends to zero in the limits of $\theta^* \to 0, \pi/2$ 
which counters the divergence in $f_3$.
Since Eq.~\eqref{eq:eta_relationship} has the same form for both sum 
and product-separable 
potentials, we may neglect the term $f_3 \eta_{\sigma s}^*$ in both cases.

A final simplification follows by noting that $f_2 = 1 + f$.
We arrive at the extremely simple approximate expressions for $\fnl$ 
\be
\begin{array}{l l l l}
\displaystyle{\frac{6}{5} \fnl} &\simeq& \displaystyle{f\, \Big[- \eta_{ss}^* +
2 \Omega \, (\eta_{ss} - \ep) \Big]} & ~~~~~~\mbox{-- Sum separable}
\,,\vspace{2mm}\\
\displaystyle{\frac{6}{5} \fnl} &\simeq& \displaystyle{ 
f\, \Big[- \eta_{ss}^* + 2 \eta_{ss} \Big]} & ~~~~~~\mbox{-- Product separable} \,,
\label{eq:fnl_direction}
\end{array}
\ee
where $f$ is positive definite. These simpler expressions 
make it transparent that the condition for $\fnl$ to be large is that $f \gg1$.
We emphasise that these approximate expressions will be extremely accurate when $\fnl>1$.

\subsection{Trispectrum}

A similar analysis can be performed for the $\tnl$ and $\gnl$ parameters. For
sum-separable potentials we find
\bea
\tnl &=&
\tau_1 {\eta_{ss}^* }^2
- \tau_2 \eta_{ss}^* \eta_{\sigma s}^*
+ \tau_3 {\eta_{\sigma s}^*}^2 
- \tau_4 \ep^* \eta_{ss}^*
+ \tau_5 \ep^* \eta_{\sigma s}^* 
+ \tau_6 {\ep^*}^2 \nonumber \\
&& \quad 
- \tau_7 \, \Omega \, \eta_{ss}^* (\eta_{ss}-\ep)
- \tau_8 \, \Omega \, \eta_{\sigma s}^* (\eta_{ss}-\ep)
+ \tau_9 \, \Omega \, \ep^* (\eta_{ss}-\ep)
+ 4 \tau \, \Omega^2 \, (\eta_{ss}-\ep)^2 \,, \qquad \quad
\label{eq:ss_tnl_full}\\
\frac{54}{50} \gnl &=&
\tau_1 {\eta_{ss}^* }^2
- \tau_2 \eta_{ss}^* \eta_{\sigma s}^*
+ \tau_3 {\eta_{\sigma s}^*}^2 
- \frac{1}{4} \tau_4 \ep^* \eta_{ss}^*
+ \frac{1}{4} \tau_5 \ep^* \eta_{\sigma s}^* 
+\frac{1}{4} \tau_2 {\xi_{sss}^*}^2
-\frac{1}{2} \tau_3 {\xi_{\sigma ss}^*}^2 \nonumber \\
&& \quad 
- \frac{3}{4} \tau_7 \, \Omega \, \eta_{ss}^* (\eta_{ss}-\ep)
- \frac{3}{4} \tau_8 \, \Omega \, \eta_{\sigma s}^* (\eta_{ss}-\ep) \nonumber \\
&&\quad +g_1 \, \Omega^{3/2} \left( \xi_{sss}^2 - 2 \eta_{\sigma s} (\eta_{ss} + \ep) \right) 
+4 g_3 \, \Omega \, \frac{W}{W^*} \cos 2 \theta \, \eta_{ss} (\eta_{ss}-\ep) \,,
\label{eq:ss_gnl_full}
\eea
and for the product-separable potentials we find
\bea
\tnl &=&
\tau_1 {\eta_{ss}^* }^2
- \tau_2 \eta_{ss}^* \eta_{\sigma s}^*
+ \tau_3 {\eta_{\sigma s}^*}^2 
- \tau_7 \eta_{ss}^* \eta_{ss}
- \tau_8 \eta_{\sigma s}^* \eta_{ss}
+ 4 \tau \eta_{ss}^2  \,,
\label{eq:ps_tnl_full}\\
\frac{54}{50} \gnl &=&
\tau_1 {\eta_{ss}^* }^2
- \tau_2 \eta_{ss}^* \eta_{\sigma s}^*
+ \tau_3 {\eta_{\sigma s}^*}^2 
+ \tau_3 \ep^* \eta_{ss}^*
+\frac{1}{4} \tau_2 {\xi_{sss}^*}^2
-\frac{1}{2} \tau_3 {\xi_{\sigma ss}^*}^2 \nonumber \\
&& \quad 
- \frac{3}{4} \tau_7 \eta_{ss}^* \eta_{ss}
- \frac{3}{4} \tau_8 \eta_{\sigma s}^* \eta_{ss}
+g_1 \left( \xi_{sss}^2 - 2 \eta_{\sigma s} \eta_{ss}\right) 
+4 g_2 \eta_{ss}^2 \,,
\label{eq:ps_gnl_full}
\eea
where the various functions occurring in these expressions are defined as
\be
\begin{array}{lllllll}
\tau_1 &=& \displaystyle{\frac{1}{\Lambda^3}(\cos ^8 \alpha \sin ^6 \theta^*+\sin ^8 \alpha \cos ^6 \theta^*)}\,, &~~~~&
\tau_4 &=& 2 f_1 f_2 \vspace{2mm}\,,  \\

\tau_2 &=& \displaystyle{\frac{\sin 2 \theta^*}{\Lambda^3}(\cos ^8 \alpha \sin ^4 \theta^*-\sin ^8 \alpha \cos ^4 \theta^*)}\,,  &&
\tau_5 &=& 2 f_1 f_3\vspace{2mm}\,,  \\

\tau_3 &=& \displaystyle{\frac{f_1}{2 \Lambda^2}(\cos ^8 \alpha \sin ^2 \theta^*+\sin ^8 \alpha \cos ^2 \theta^*)}\,,  &&
\tau_6 &=& f_1^2\vspace{2mm}\,,  \\

\tau_7 &=& \displaystyle{\frac{4 f}{\Lambda} \left(\cos^2 \alpha \sin^2 \theta^* + \sin^2 \alpha \cos^2 \theta^*\right)}\,,  &&
\tau_9 &=& 4 f_1 f \vspace{2mm}\,,  \\

\tau_8 &=& \displaystyle{-\frac{ \sin 2 \theta^* \sin^2 2 \alpha}{2\Lambda^2} \left(\cos^2 \alpha - \cos^2 \theta^* \right)}\,,  &&
\tau &=& \displaystyle{\frac{f \sin^2 2 \alpha}{4 \Lambda}}\vspace{2mm}\,,  \\

g_1 &=& \displaystyle{g_3\sin 2 \alpha }\,, \qquad\qquad
g_2 = \displaystyle{g_3\cos 2 \alpha }\,, &&
g_3 &=& \displaystyle{-\frac{f}{2 \Lambda}(\cos^2 \alpha - \cos^2 \theta^*)}\,.
\label{eq:ss_tnl_para}
\end{array}
\ee

We can now proceed similarly as we did for the bispectrum and show that a number 
of these terms are negligible if the trispectrum parameters are large.
Since the trispectrum functions pre-multiply quantities that are second order in slow-roll, 
in this case we neglect any functions that are never larger than $10$, rather than order of unity.
The details of this analysis can be found in appendix \ref{appendixB:fullcalcs}.

\subsubsection{$\tnl$} 

For $\tnl$, we find the remarkably simple forms
\be 
\begin{array}{l l l l}
\displaystyle{\tnl} &\simeq& \displaystyle{{\cal C} \left( \frac{6}{5} \fnl
\right)^2 - \frac{12}{5} \fnl (\eta_{ss}^* - f_1 \ep^*)} & ~~~~~~\mbox{-- Sum
separable} \,,\vspace{2mm}\\
\displaystyle{\tnl} &\simeq& \displaystyle{{\cal C} \left( \frac{6}{5} \fnl
\right)^2 - \frac{12}{5} \fnl \eta_{ss}^*} & ~~~~~~\mbox{-- Product separable} \,,
\label{eq:tnl_pre_final}
\end{array}
\ee
where
\be
{\cal C} = \frac{\tau}{f^2} = \frac{\Lambda}{(\cos^2 \alpha - \cos ^2 \theta^*)^2} \,.
\ee
A further simplification can be made by noting that  
${\cal C} \geq 1$ and that $f_1$ is at most of order unity.
The second terms will therefore be suppressed relative to the first by 
at least ${\cal O}(\ep^*)$. We thus find
\be
\tnl \simeq {\cal C} \left( \frac{6}{5} \fnl \right)^2 \,,
\label{eq:tnl}
\ee
which is valid for both sum and product-separable potentials.

In ref.~\cite{Smidt:2010ra}, the ratio between $\tnl$ and $\left( \frac{6}{5} \fnl \right)^2$ 
was parameterised by $A_\mathrm{NL}$. Peterson {\it et al.} subsequently showed that 
$A_\mathrm{NL} = 1/r_c^2$ for two-field models under slow-roll \cite{Peterson:2010mv}, 
where $r_c$ determines the fraction of the curvature perturbation which is sourced by
a horizon crossing isocurvature mode.
Our result, Eq. \eqref{eq:tnl}, is complementary to this analysis, explicitly showing the form of 
$r_c$ in terms of the dynamics of inflation for separable potentials.

\subsubsection{$\gnl$} 

Unsurprisingly, $\gnl$ does not simplify so neatly.
In the product-separable case we find
\bea
\frac{54}{50} \gnl &\simeq& 
\tnl \left( \frac{\eta_{ss}^* -\eta_{ss}}{\eta_{ss}^* -2\eta_{ss}} \right) 
-\frac{6}{5} \fnl (2 \eta_{ss}^* + \eta_{ss})-g_4 {\xi_{sss}^*}^2 
+ g_1 \Big[\xi_{sss}^2 - 2 \eta_{\sigma s} \eta_{ss}\Big] \,,
\label{eq:ps_gnl}
\eea
where $g_4 = \frac{1}{4} \left(\tau_3 \sin 2 \theta^* \cos 2 \theta^* -\tau_2 \right)$. 
For sum-separable potentials we obtain two additional terms, and the expression takes the form
\bea
\frac{54}{50} \gnl &\simeq& 
\tnl \left( \frac{\eta_{ss}^* -\Omega \, (\eta_{ss}-\ep)}{\eta_{ss}^* -2 \, \Omega \, (\eta_{ss}-\ep)} \right) 
-\frac{6}{5} \fnl (2 \eta_{ss}^* + \Omega \, (\eta_{ss}-\ep))
-g_4 {\xi_{sss}^*}^2 \nonumber \\
&& + g_1\, \Omega^{3/2} \, \Big[\xi_{sss}^2 - 2 \eta_{\sigma s}
(\eta_{ss}+\ep)\Big] -\frac{1}{2} f_1 f \ep^* \eta_{ss}^* \nonumber \\
&& + 4 g_3 \, \Omega \, (\eta_{ss}-\ep) \left( \frac{W}{W^*}\cos 2 \theta
\eta_{ss} - \Omega \cos 2 \alpha (\eta_{ss} - \ep) \right)
\,.
\label{eq:ss_gnl}
\eea

These formulae for $\gnl$ are once again complimentary to 
those derived by Peterson {\it et al.} \cite{Peterson:2010mv}. 
One advantage of our results is that because of their explicit nature, we can easily consider which
shapes in the inflationary potential will cause the different terms 
in Eqs. \eqref{eq:ps_gnl} and \eqref{eq:ss_gnl} to become large. 
Furthermore,  the
Horizon Crossing Approximation is easily implemented in our formulae by 
taking $\Omega\to0$.

\section{Analysis}

We now turn to the interpretation of the formulae we have presented in the previous section. 

\subsection{Bispectrum}
\label{sec:bispectrum_analysis}

\begin{figure}[htb]
\centering
\includegraphics[width=0.5\textwidth]{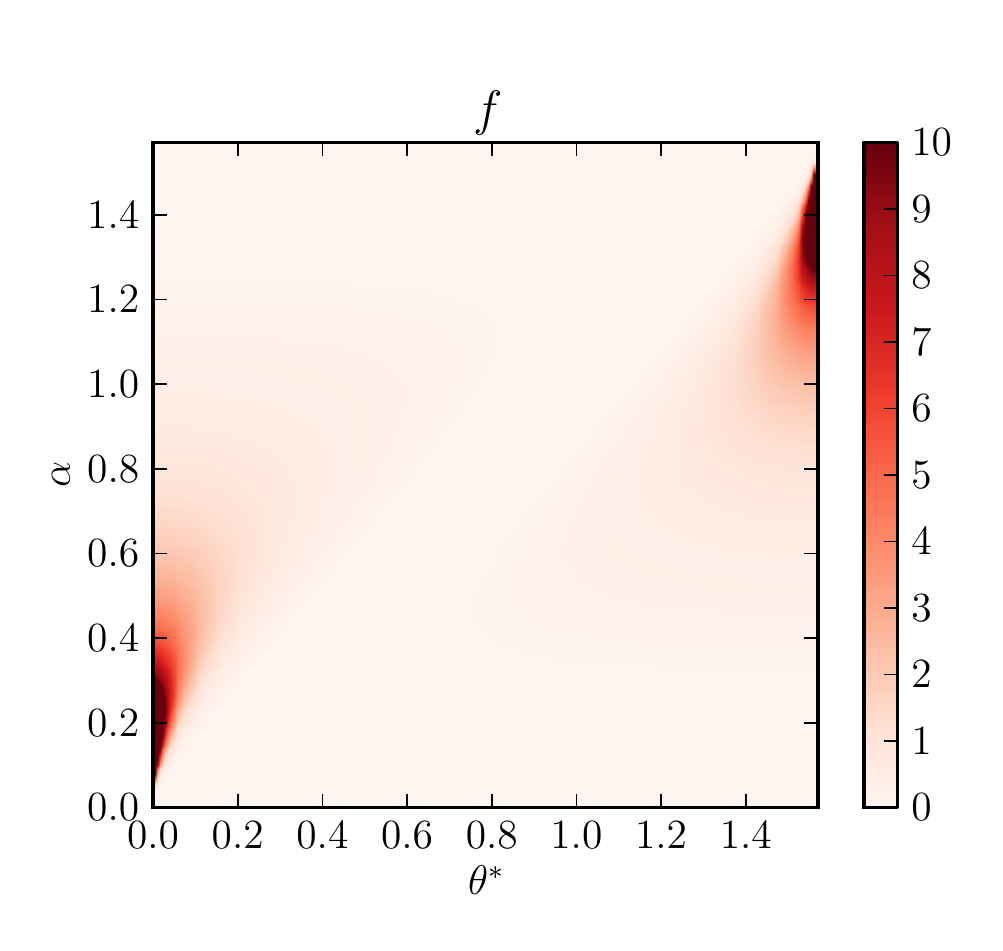}
\caption{Heatmap of the function $f$. Since $\fnl$, $\eta_{ss}$ and $\ep$ are 
symmetric under field exchange $\phi \leftrightarrow \chi$, the function $f$ must have the same symmetry. We can see this by inverting the heatmap though the point 
$(\theta^*,\alpha) = (\pi/4 , \pi/4)$ which leaves $f$ invariant.}
\label{fig:f}
\end{figure}

The expressions \eqref{eq:fnl_direction} imply that a necessary condition for $|\fnl| > 1$ is that 
$f \gg 1$. We can see when this occurs by plotting $f(\alpha, \theta^*)$, as
shown in Fig. \ref{fig:f}.

For product-separable potentials, $\alpha$ and $\theta$ are trivially 
related as $\alpha = \theta$. The initial (horizon crossing) value of
$\alpha$ for product-separable potentials is therefore $\alpha_{\rm init} = \theta^*$.
For sum-separable potentials the same initial condition is found. This can be seen by 
taking the limit $c \to *$ in Eq.~\eqref{eq:ss_uv}.
This means that all separable inflationary models start on the diagonal of the heatmaps at 
horizon crossing. Since $f=0$ on the diagonal, we know that initially $\fnl$ 
will be given by the various negligable terms that we have dropped in \eqref{eq:fnl_direction}, and so $|\fnl| \ll 1$ initially, as we expect \cite{Seery:2005gb}.

As the model evolves from a given $\theta^*$, $\alpha$ varies such that
the model traces a vertical line on the heatmap. 
For this trajectory to ever intercept one of the regions in which $f \gg 1$
the initial conditions must be such that the initial phase-space velocity is
dominated by one of the two fields $\phi$ or $\chi$.

\subsubsection{The product-separable case}

As an example, let us take a product-separable model with $\theta^* \ll 1$,
such that the horizon crossing conditions correspond to a position on the 
diagonal of the heatmap of $f$ in the lower left-hand corner.
If the angle $\theta$ ($=\alpha$) increases towards $\pi/2$, there
can be a transient `spike' in $\fnl$ as the model passes through the region in which $f$ 
is large. If the trajectory turns back, so that $\theta$ decreases, it may well pass 
back through this region again and another transient signal in $\fnl$ can be produced. 
Of course, whether or not a significant spike will occur is also dependent on the 
slow-roll parameters that define the model. 
The magnitude of this spike in $\fnl$ increases as $\theta^*$ decreases towards zero.
We must be careful with the interpretation, however, since our analysis presumes separable
potentials and this places strong constraints on the possible modes of behaviour. 
If the initial field velocity is \emph{exactly} aligned to one of the field axes then it will
(classically) remain so indefinitely, since this implies $\theta = \theta^*$ 
\emph{always}. Furthermore, for neighbouring initial conditions with sufficiently small values  
of $\theta^* $, one may find that $\theta$ will take longer than the required length of inflation 
(roughly 60 efolds) to grow sufficiently for there to be 
a significant enhancement of $\fnl$. There will therefore be 
an upper bound on the value $\fnl$ achievable by any such 
potential\footnote{There will also be another upper bound on $|\fnl|$ imposed 
by the quantum scatter of the field near horizon crossing which 
will prohibit the limit $\theta^* \to 0$ being physically realisable.}.

\subsubsection{The sum-separable case}

To make progress with understanding the dynamics in the sum-separable case,
it is necessary to understand the relationship between $\alpha$ and $\theta$.
To do this we differentiate Eq.~\eqref{eq:alpha} to find 
\be
\alpha' = \frac{W}{W^*} \frac{\sin^2 2 \theta}{\sin^2 2 \alpha} \theta' \,,
\label{eq:alphadash}
\ee
where a prime represents a derivative with respect to some time variable.
The fractions in Eq.~\eqref{eq:alphadash} are positive definite and so $\alpha$
increases as $\theta$ increases and vice versa.

\paragraph{Vacuum domination.} More progress is possible by considering 
the vacuum-dominated limit, where $W \simeq W^*$. This is a 
good approximation for some models of inflation such as hybrid inflation. 
Eliminating the ratio $W/W^*$ from 
Eq.~\eqref{eq:alphadash},  we see that $\alpha \simeq
\theta$. Moreover, one also finds $\Omega \simeq 1$ in this limit.
Consequently, for vacuum-dominated sum-separable potentials, one 
may use the heatmaps in the same way as for product-separable potentials.
Furthermore, we see that in this case the sum and product-separable formulae 
for $\fnl$ are identical, apart from the presence of the slow-roll parameter $\ep$ in the
sum-separable case. 
Thus for models with $\eta_{ss} \gg \ep$ there is a very near equivalence
between product-separable potentials 
and vacuum-dominated sum-separable potentials in terms of their contribution to $\fnl$.

\paragraph{General models.} We now consider Eq.~\eqref{eq:alphadash} 
for general sum-separable models without vacuum-domination.
The ratio $W/W^*$ is initially unity and decreases towards zero. This ensures that, 
whilst the angle is monotonically varying, $\alpha$ lags behind $\theta$.
Furthermore, the difference between $\alpha$ and $\theta$ will become 
more pronounced the smaller the ratio $W/W^*$. When this ratio goes to zero, we see that 
$\alpha$ remains constant despite any subsequent turning of the trajectory in phase space.
Similarly, $\alpha$ will cease to evolve if either of the limits $\theta \to 0$ or 
$\theta \to \pi/2$ are reached. Physically, these limits correspond to straight lines in 
the field phase space under which conditions it is well known that $\zeta$ does not evolve. 
We note that $\Omega$ is zero in any of these three limits and so we are only left with 
the $\eta_{ss}^*$ term in $\fnl$. This is the approximate HCA formula for $\fnl$. 
We note that since $\alpha \neq \theta$, it is quite possible for $f$ to be large when 
$\alpha$ becomes a constant, and so produce an observationally relevant  constant $\fnl$. 
For a given evolution, the final value of $\alpha$ is readily extracted once the initial and final 
field values are known, and hence one can check whether the correct value of $\alpha$ is  
reached in order to give a large non-Gaussianity.

The final case to consider is when the trajectory turns back on itself during
its evolution.  There is no barrier to constructing sum-separable potentials 
which exhibit this behaviour for particular evolutions.
This will mean that the model moves up and then down a vertical line on 
the heatmap, perhaps many times.
To fully quantitatively understand how this movement occurs requires a knowledge of how the potential 
$W$ varies with $\theta$, which is necessarily model-specific. 
However, it is possible to gain some more detailed intuition for how $\alpha$ 
varies with $\theta$ by rewriting Eq.~\eqref{eq:alphadash} as
\be
h'(\alpha) = \frac{W}{W^*} h'(\theta)\,,
\label{eq:alphadash2}
\ee
where we have defined the function $h(x) = 4x - \sin 4x$, which is monotonic in $x$.
One sees that $h(\alpha)$ and $h(\theta)$ increase and decrease simultaneously and hence the
velocity $h'(\alpha)$ is always smaller than the velocity $h'(\theta)$. Since $W/W^*$
is constantly decreasing then so is the range of values of $h(\alpha)$ which the evolution can 
reach. Ultimately $h'(\alpha) \to 0$ and $h(\alpha)$ takes a constant value.
Since $h(\alpha)$ is monotonic in $\alpha$ we see that restricting the range of $h(\alpha)$ 
translates into introducing `excluded regions' at the top and bottom of the heatmaps,
the size of which will grow as the potential drops, and ultimately the whole of the heatmap will 
be excluded except for the final value of $\alpha$. Once again, for a given evolution the final 
value of $\alpha$ can be readily calculated, and one can check if it is in the observationally 
relevant regime.

\subsubsection{The role of $\Omega$}

The value of $\Omega$ determines which of the terms $\eta_{ss}^*$ or 
$\eta_{ss}$ dominates in Eq.~\eqref{eq:fnl_direction}.
For product-separable or vacuum-dominated sum-separable potentials, 
$\Omega$ is effectively fixed to be unity, and so which term dominates 
depends on how $\eta_{ss}$ evolves during the evolution. On the other hand, 
if $\Omega \to 0$ then the $\eta_{ss}^*$ term will dominate. 
It is instructive, therefore, to think of inflationary models belonging to one of two classes, those for which $\Omega \simeq 1$ throughout
and those for which $\Omega \to 0$ at some point.
We now briefly consider each of these cases in turn.

\paragraph{Evolutions for which $\Omega \simeq 1$.} These models either 
have product-separable or vacuum-dominated sum-separable potentials.
In the simplest cases of interest, such as falling from a potential ridge or rolling into a vacuum 
dominated valley \cite{Elliston:2011dr}, the absolute value of the potential does not 
change significantly during this phase of evolution.
If $\eta_{ss} \simeq \eta_{ss}^*$ then one finds $\frac{6}{5}\fnl \simeq f \eta_{ss}^*$. 
Since $f$ is positive definite, we see that the sign of $\fnl$ is the same as the sign of the
 isocurvature mass. This is in agreement with previous calculations \cite{Elliston:2011dr}
where a ridge shape ($\eta_{ss} < 0$) was shown to lead to a negative $\fnl$ 
and a valley shape ($\eta_{ss} > 0$) was shown to lead to a positive $\fnl$.
If inflation does not end abruptly, for example by a hybrid transition, but the fields continue 
their evolution, then $\fnl$ will continue to evolve until $\theta \to 0$ or $\theta \to \pi/2$. 
In either of these limits $f\to 0$ and so $\fnl$ is much smaller than unity.

\paragraph{Evolutions for which $\Omega \to 0$.} These models naturally reach a limit in which 
$\alpha$ becomes constant, and the HCA becomes a good approximation. This may or may not coincide with a scenario in which the
isocurvature perturbations decay and an adiabatic limit is reached.
If an adiabatic limit is arrived at, then we can be certain that there will be no 
further evolution of $\zeta$ and its statistics.
When $\Omega \to 0$ we find $\frac{6}{5}\fnl \simeq -f \eta_{ss}^*$. 
Due to the minus sign, if we begin in a region with a negative isocurvature mass, 
for example on a ridge, then such a model will eventually produce a positive $\fnl$, 
whereas if we begin in region with a positive isocurvature mass, then $\fnl$ will ultimately 
reach a negative limiting value. \\

In summary, the positive $\fnl$ that is marginally preferred by the WMAP 
data \cite{Komatsu:2010fb} can be generated in two possible ways. Either, the evolution must 
begin with a small $\theta^*$ (or one close to $\pi/2$) and an initially 
\emph{negative} isocurvature mass, and then evolve until $\alpha$ naturally takes a small 
constant value suitable to make $f$ large. 
Or alternatively, the evolution must evolve from a small $\theta^*$ (or one
close to $\pi/2$) 
during which time the isocurvature mass is 
\emph{positive}, and some mechanism much interrupt the dynamics whilst $f$ is large.

\subsection{Trispectrum}\label{sec:trispectrum_analysis}

We now turn to analyse the trispectrum. Our aims are twofold. First,
to understand the types of models and the initial and final conditions for which the 
observational non-Gaussianity parameters $\tnl$ and $\gnl$ can be large.
Secondly, to infer if it is possible to relate these non-Gaussianity 
parameters to one another, perhaps for 
specific classes of models. Such relations are potentially important in order to 
observationally exclude classes of models.

\begin{figure}[htb]
\begin{minipage}{0.5\linewidth}
\centering
\includegraphics[width=\textwidth]{f.pdf}
\end{minipage}
\begin{minipage}{0.5\linewidth}
\centering
\includegraphics[width=\textwidth]{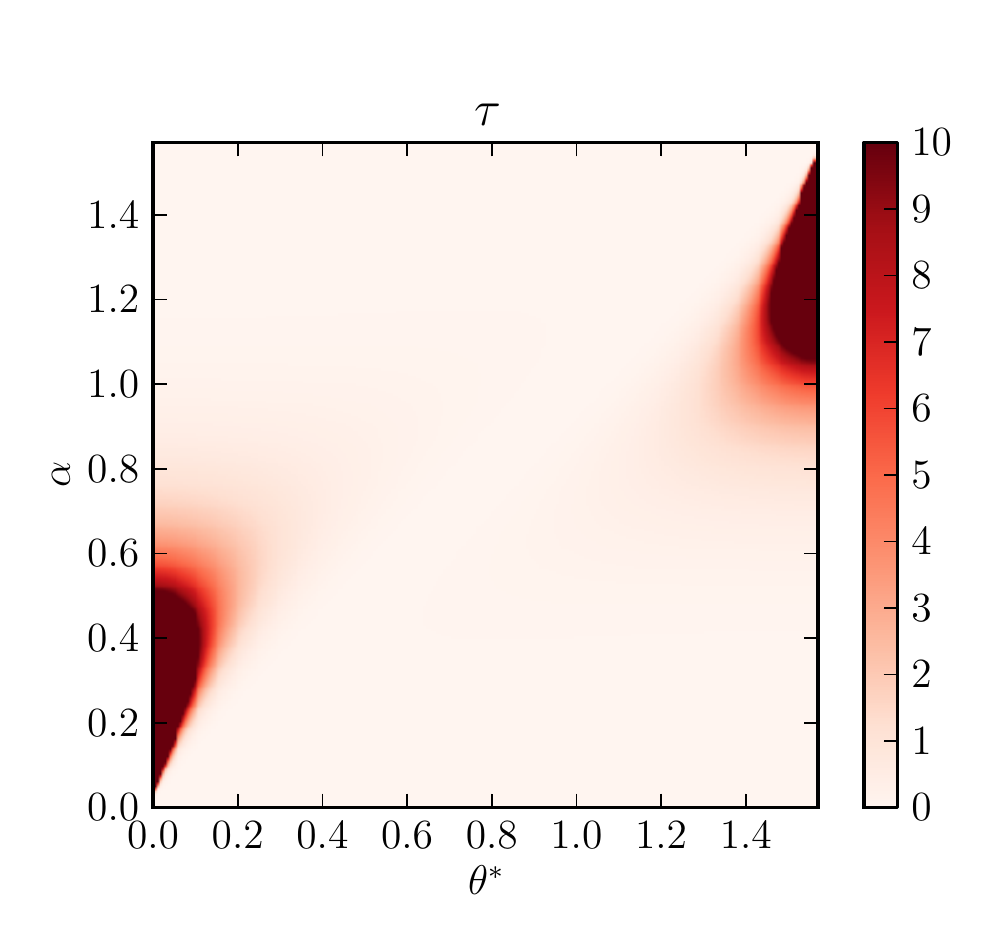}
\end{minipage}
\caption{Heatmaps of $f$ and $\tau$ on the same scale. We note that both are large in the same regions of the parameter space.}
\label{fig:ft}
\end{figure}

Referring to Eqs. \eqref{eq:tnl_pre_final} we see that $\tnl$ can only be large 
when $\tau$ or $f$ are large. We plot these functions side-by-side 
in Fig.~\ref{fig:ft}. The fact that both functions peak in similar regions of
the parameter space means that the classes of models that are capable of producing 
a large $\tnl$ are the same as the classes of models that can produce a large $\fnl$. 
%
\subsubsection{$\tnl$ and $\fnl$ relation}
\begin{figure}[htb]
\centering
\includegraphics[width=0.5\textwidth]{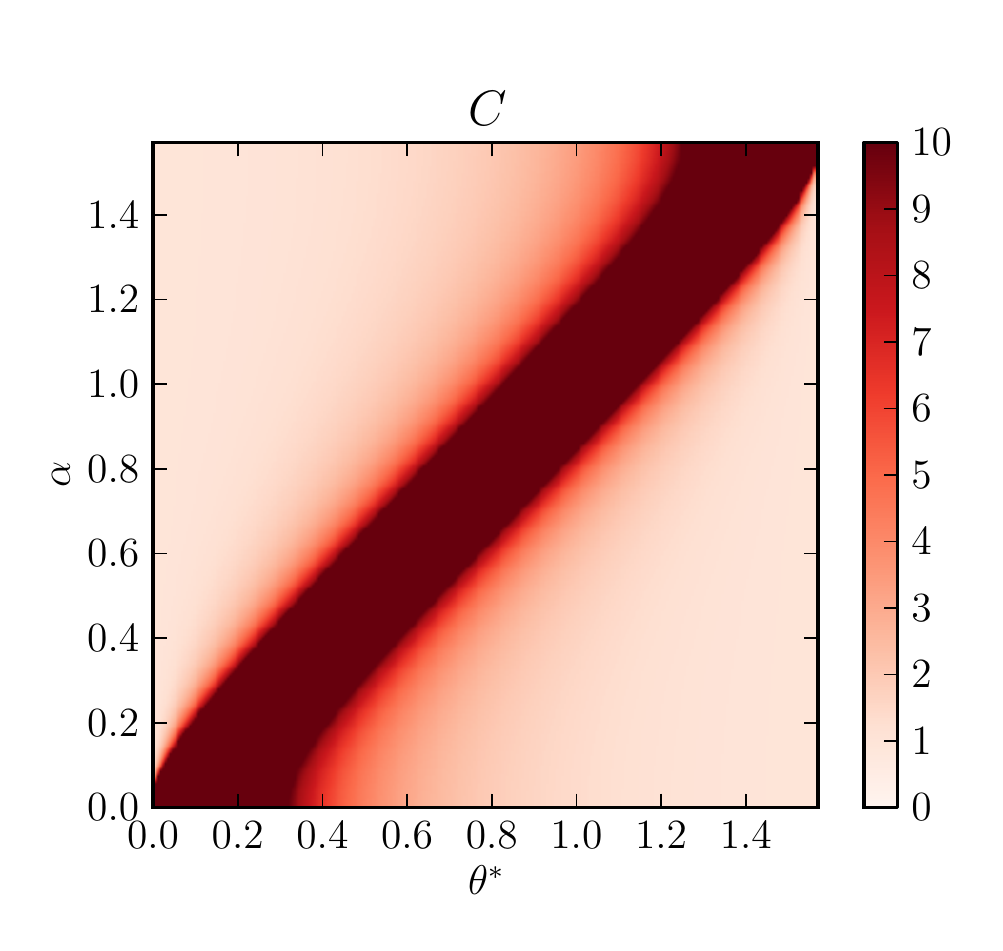}
\caption{Heatmap of the function $\C$ illustrating the conditions required for a model to deviate from the single field consistency result for $\tnl$.
This is only physically interesting in the regions for which $f$ is  large which are close to the sides of the heatmap.
The region where $\C\gg1$ overlaps with part---but not all of---the region where $f$ is large
and one can see that this overlapping region is displaced from the sides of the heatmap.}
\label{fig:C}
\end{figure}

For canonical scalar field inflation, $\tnl$ and $\fnl$ satisfy the condition \cite{Suyama:2007bg}
\be
\tnl \geq \left( \frac{6}{5} \fnl \right)^2 \,,
\label{eq:sf_consistency}
\ee
where equality occurs for single field inflation\footnote{See also the recent 
work by Sugiyama \cite{Sugiyama:2012tr} claiming that this equality is broken 
when calculations contributions are included from all loops.}.
This is fully consistent with our result \eqref{eq:tnl}, once we recall that $\C\geq1$.
We plot $\C$ in Fig.~\ref{fig:C}. The interesting regions of this plot are those for 
which $\tnl$ is not small and so $f^2 \gg 1$ as shown in Fig.~\ref{fig:ft}.
We see that if a model has $\theta^* \ll 1$ and $\theta$ subsequently increases, then 
such a model will first enter the region in which $\C\gg1$
and so $\tnl$ will grow whilst $\fnl$ remains small. 
It is unsurprising that $\tnl$ evolves first, since being associated with a higher order moment,
it will be more sensitive to outliers of the $\delta N$ distribution.
For large $\alpha$, one can see that $\C \simeq 1$ to a very 
good degree of accuracy, and the single field relation becomes a good approximation.
We note that $\tnl$ deviates more from the single-field limit $\C = 1$ 
when $\theta^*$ is fine-tuned to be closer to the  $\theta^*=0$ axis.

\subsubsection{$\gnl$ and $\tnl$ relations}
\label{sec:gnltnl}

We now turn our attention to consider possible relations between $\gnl$ and $\tnl$.

\paragraph{The $\Omega=0$ case.}  First, we consider the case of non-vacuum 
dominated sum-separable potentials for evolutions which reach $\Omega=0$ such that
all of the observational parameters have ceased to evolve. From Eq.~\eqref{eq:ss_gnl} we
see that 
\bea
\frac{54}{50} \gnl &\simeq& \tnl - \frac{3}{5} \fnl \left(4\eta_{ss}^*- f_1
\ep^* \right) - g_4 {\xi_{sss}^*}^2 \,.
\label{eq:tnl_gnl_full}
\eea
The $\fnl$ term in Eq.~\eqref{eq:tnl_gnl_full} is suppressed relative to the
$\tnl$ term by a relative factor of $\fnl^{-1} \times {\cal O} (\ep^*)$ and so
may be safely neglected. 
In cases were  ${\xi_{sss}^*}^2$ can be neglected, for example in the absence of
significant terms in the potential beyond quadratic order at horizon crossing, we
find a relation between $\tnl$ and $\gnl$ as
\be
\frac{54}{50} \gnl \simeq \tnl \,.
\label{eq:gnl_tnl}
\ee

\begin{figure}[htb]
\centering
\includegraphics[width=0.5\textwidth]{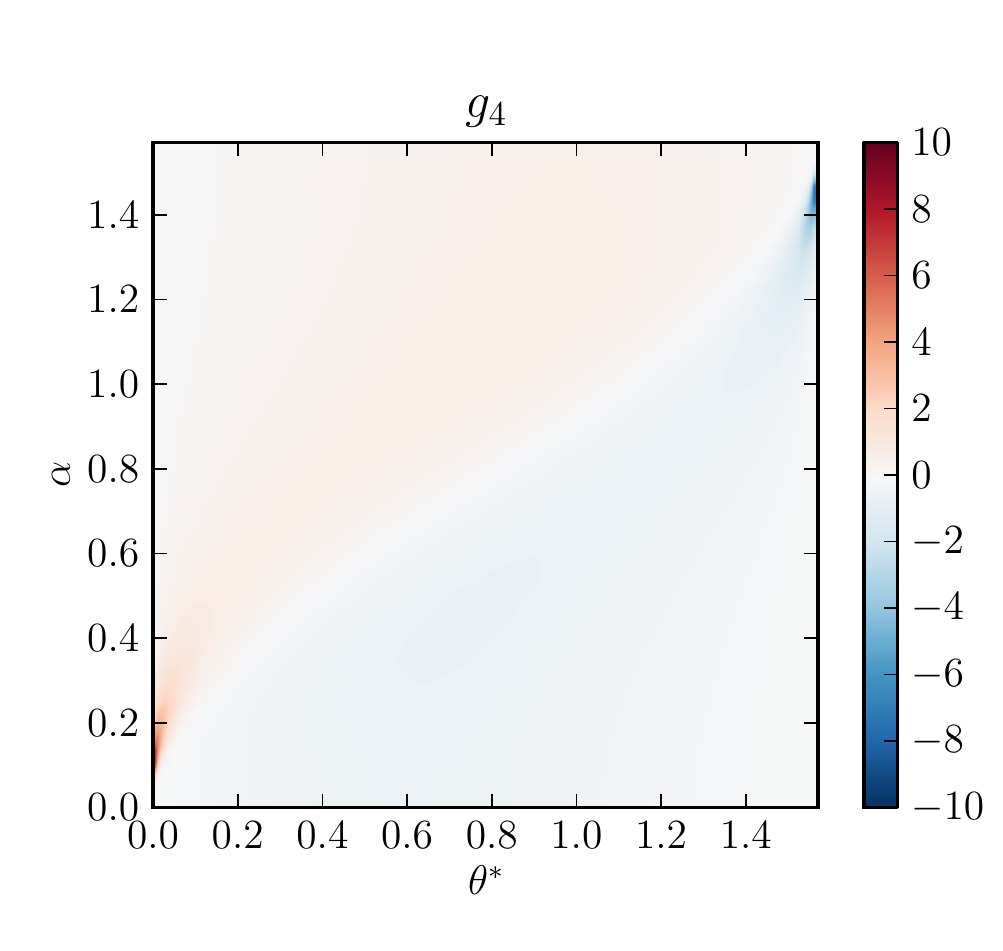} 
\caption{Heatmap of the function $g_4$ which is appropriately antisymmetric
about field exchange. The very small lobe in the bottom left hand corner has positive $g_4$, 
whilst the lobe in the top right hand corner has negative $g_4$.
This pattern is repeated for $g_1$ and $g_3$ as seen in Fig.~\ref{fig:gg}.}
\label{fig:g4}
\end{figure}

Next, let us look at the term $g_4 {\xi_{sss}^*}^2$ and assess when it may be
relevant. If an inflationary potential exists for which this term is important then
it will have a signature such that $\gnl$ deviates from Eq. \eqref{eq:gnl_tnl}.
The heatmap for $g_4$ is plotted in Fig. \ref{fig:g4} and we see that the areas in 
which $g_4$ is large are very small in comparison to the corresponding areas for the $\tau$ function.
Practically speaking, this ensures that one has to tune the parameters of the model
to a very high degree in order to access this region. 
In \S \ref{sec:transient_calcs}, we show that, in the interesting limit where 
$\theta^*$ and $\alpha$ are small, we can accurately approximate $2 g_4 \simeq \theta^* \tau$. 
Thus a necessary (but not sufficient) condition for $|\gnl|$ to be large is that
$\theta^* \tau \gg 1$. In addition, for $\gnl$ to deviate from Eq. \eqref{eq:gnl_tnl} 
we require that $L = \theta^* {\xi_{sss}^*}^2 / 2{\eta_{ss}^*}^2$ is not small. 
Potentials with $L \gg 1$ therefore have the capacity to generate $|\gnl|\gg\tnl$.

Considering a sum separable potential, in the case where $\theta^* \ll1$, 
to leading order we find
\be
{\eta_{ss}^*}^2 = {\eta_{\chi}^*}^2\,,\quad
{\xi_{sss}^*}^2 = {\xi_{\chi}^*}^2\,,\quad
\theta^* = \sqrt{\epc^*/\epp^*}\,,
\ee
where we have indicated  $\eta_{\chi \chi} = \eta_\chi$ and $\xi_{\chi \chi \chi}^2 = \xi_{\chi}^2$.
Expanding out the slow-roll parameters in terms of potential derivatives, one finds that 
$L \simeq W_{,\chi} W_{,\chi \chi \chi} / 2 W_{,\chi \chi}^2$.
If one considers a general power-law potential $V(\chi) = V_0 \chi^n$ then 
$L=0$ for $n=0,1,2$ 
and $L \leq 1/2$ for $n \geq 3$.
On the other hand, for an exponential potential 
$V(\chi) = V_0 e^{\lambda \chi}$ then $L=1/2$. 
In these two cases we would therefore not expect to find a deviation from 
Eq.~\eqref{eq:gnl_tnl} beyond a factor of 2. It is interesting to note that a
potential 
of the form $V(\chi) = a \ln (\chi-b)$ for constants $a$ and $b$ has $L=1$ and so for 
such a potential the two leading order terms in $\gnl$ exactly cancel and so $\gnl$ 
is of order $\fnl \times {\cal O} (\ep^*) $.

We now ask if there is a potential with $L \gg 1$. Considering 
a polynomial potential  $V(\chi)$, the necessary condition is for the potential
to possess a linear term, a negligible quadratic term
and at least one term beyond quadratic order. 
The simplest such potential is a sloping inflection point of the form
$V(\chi) = V_0 + h \chi + \frac{1}{6} \lambda \chi^3$.
Nearby the inflection point one has $\eta_\chi \simeq 0$ whilst 
$\theta \, \xi_\chi^2 = \Mpl^3 h \lambda / W_0^2$ and so $L$ diverges. 
We shall examine such an inflection point further in \S \ref{sec:models}.

\paragraph{The $\Omega \neq 0$ case: }

\begin{figure}[htb]
\begin{minipage}{0.5\linewidth}
\centering
\includegraphics[width=\textwidth]{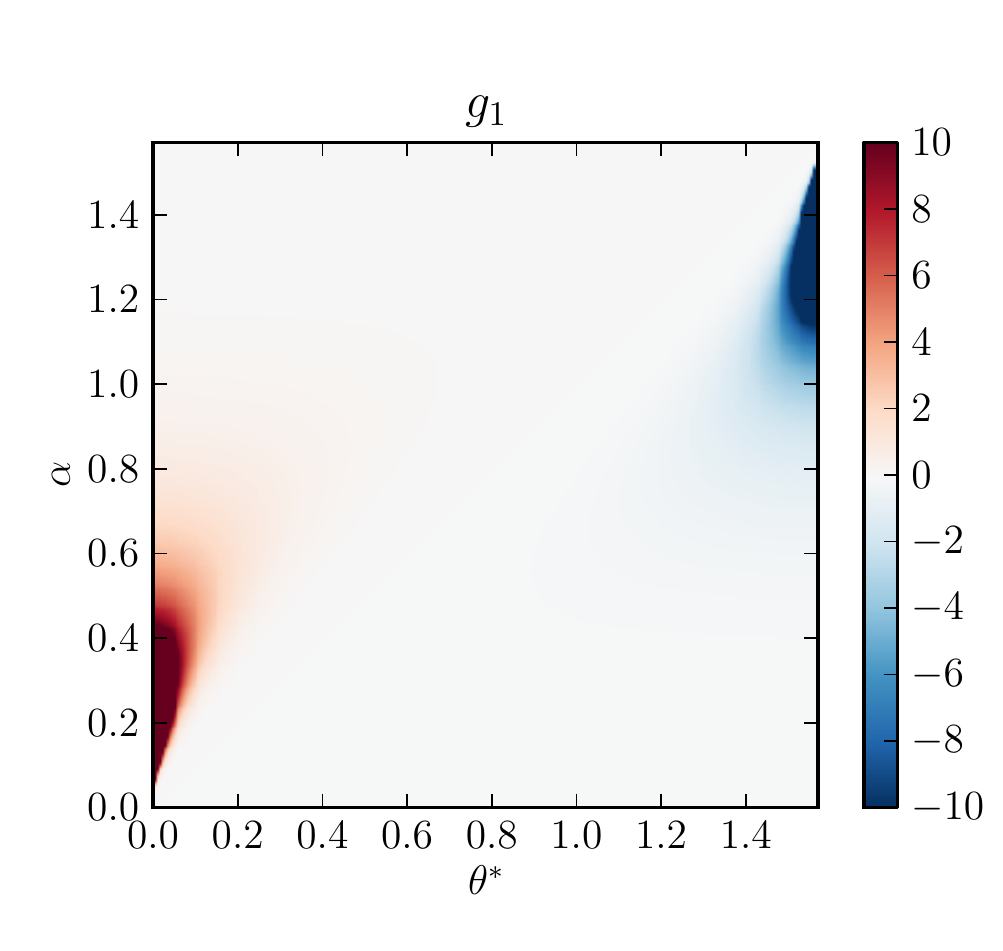}
\end{minipage}
\begin{minipage}{0.5\linewidth}
\centering
\includegraphics[width=\textwidth]{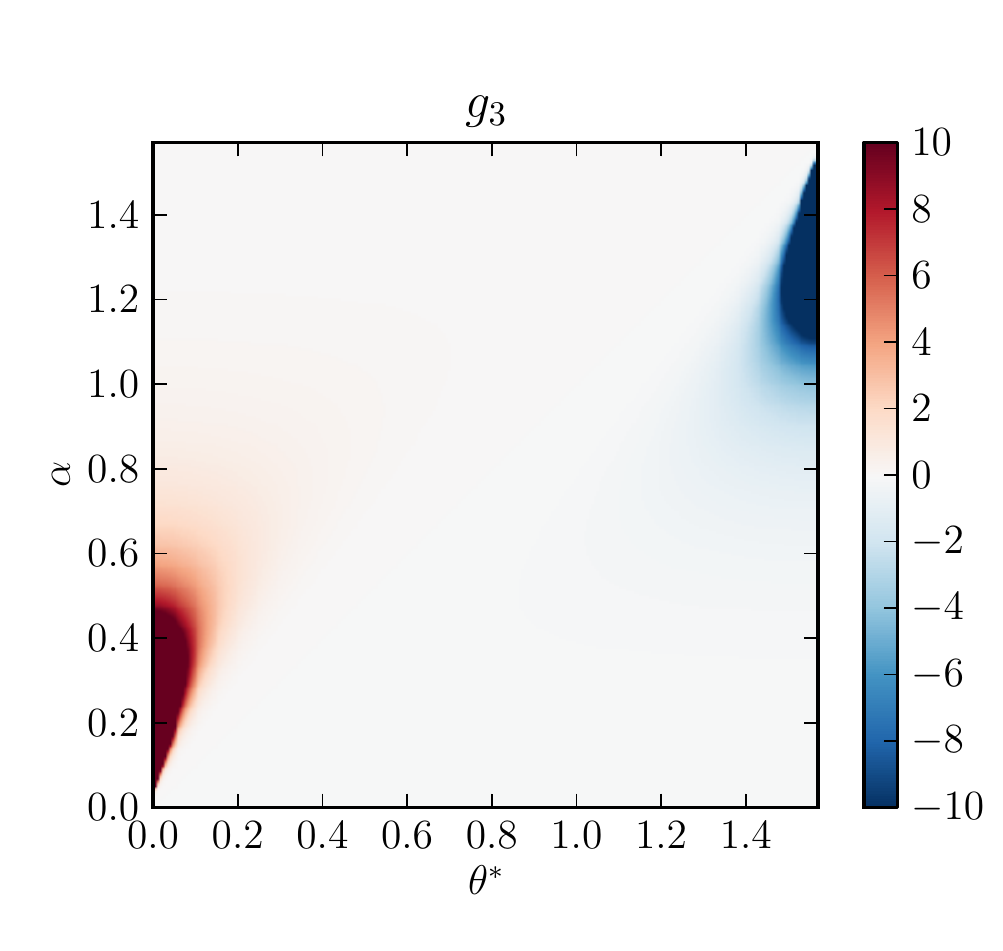}
\end{minipage}
\caption{Plots of $g_1$ and $g_3$
on the same scale as used for the other heatmaps. We note that they are very similar 
and are anti-symmetric under field interchange. Both are large in the same regions of the parameter space as $\tau$.}
\label{fig:gg}
\end{figure}

It is considerably harder to make concrete statements about the value of the 
trispectrum parameter $\gnl$, and its relation to $\tnl$, when $\Omega \neq 0$. 
We can take a step in this direction by plotting the heatmaps for the remaining functions 
which appear in the expressions \eqref{eq:ps_gnl} and \eqref{eq:ss_gnl}, $g_1$ 
and $g_3$. These are shown in Fig.~\ref{fig:gg}. 
The similarity between these plots, and those of $f$ and $\tau$, tells us that the types of 
inflationary potential, and initial conditions, that can give rise to a large 
$\gnl$ are similar to those that may give rise to a large $\fnl$ and $\tnl$. No new regions 
of interest appear for $\gnl$ which were not present before.

\paragraph{The $\Omega \simeq 1$ case: } Finally, we  consider the 
case $\Omega \simeq 1$. This arises for the vacuum dominated potentials, and so can 
be relevant for models where inflation is terminated suddenly, perhaps through a 
waterfall transition. During a vacuum dominated phase of evolution, the fields 
evolve only very slowly, and likewise turns in field space progress slowly. 

It is instructive to consider how the non-Gaussianity parameters given by 
Eqs. \eqref{eq:fnl_direction}, \eqref{eq:tnl} and \eqref{eq:ss_gnl} may be approximated 
for $\Omega \simeq 1$, $\theta \ll 1$ and a power-law potential of the form
$V(\chi) \propto \chi^n$ where we presume $n$ to be a positive integer greater than or equal to $2$. 
In the regime of interest where $\theta \gg \theta^*$ 
(such that the functions $f$ and $\tau$ may be large) we require $|\chi| \gg |\chi^*|$.
If $n=2$, such that the potential has a parabolic shape, then $\eta_{ss}$ is approximately constant
and from Eq. \eqref{eq:ss_gnl} for $\gnl$ we see that the first term will be negligible in this case.
Furthermore, if the potential is well-described by an expansion to quadratic order we see that the other
terms in $\gnl$ are also suppressed with respect to $\fnl$ and so $\gnl$ is not large. 
This was the case found by Byrnes {\it et al.} \cite{Byrnes:2008zy}
in their study of vacuum dominated quadratic potentials. 

However, for $n \geq 3$ the situation is different. The isocurvature mass $\eta_{ss}$ is now a function of
$\chi$ and $|\eta_{ss}|$ will grow as $|\chi|$ grows. In the regime of interest 
where $|\chi| \gg |\chi^*|$ we find that $\frac{54}{50} \gnl \simeq \frac{1}{2} \tnl$ 
from the first term in Eq. \eqref{eq:ss_gnl} and so
it is clear that $\gnl$ is not generally a vanishing quantity when the potential is described by terms 
beyond quadratic order. We note that the remaining terms in Eq. \eqref{eq:ss_gnl} for $\gnl$
will mostly be small in this limit, but we will anticipate that the $g_1 \xi_{sss}^2$ term will 
contribute notably to the value of $\gnl$ during this transient evolution. We shall look at the 
specific case of $n=3$ in \S \ref{sec:inflectionpot}.

\section{Typical shapes in the inflationary potential} \label{sec:transient_calcs}

In this section we briefly show how the formulae that we have found for a 
large $\fnl$, $\tnl$ and $\gnl$ can be used to make quantitative predictions in certain 
generalised settings. We do this by studying how the bi- and tri-spectra transiently evolve 
as the field space trajectory passes over particular features in the inflationary 
potential. Motivated by the analyses in the previous sections, as well as other
work \cite{Alabidi:2006hg,Byrnes:2008zy,Elliston:2011et}, we consider 
potentials with one of the following three general features:
\begin{itemize}
\item a ridge,
\item a valley,
\item an inflection point.
\end{itemize}
Our analysis leads to simple expressions which give the sign and the peak values of 
the observable parameters generated by these features.
By Taylor expanding a real potential about the initial conditions corresponding to
a given inflationary potential, it is often possible to approximate the actual potential 
by one of these features over a range of field values. 
This can allow quantitative information about the predictions of models of inflation to be 
obtained readily without the need for detailed calculations \cite{Elliston:2011dr}.

The analyses of heatmaps show that large non-Gaussianity parameters require 
small angles in the parameter space which correspond to potentials for which the 
motion is highly aligned to one of the field directions, which we take to be $\phi$ without 
loss of generality. We can then expand the functions such as $f$ 
in the limit where $\theta^*$ and $\alpha$ are small to find
\be
f \simeq \frac{\theta^6}{(\theta^4 + \theta_*^2)^2}\,, \qquad 
\tau \simeq \frac{\theta^8}{(\theta^4 + \theta_*^2)^3}\,,
\ee
and the other parameters are related as $g_1 \simeq \theta \tau$, $2 g_3 \simeq \tau$
and $2g_4 \simeq \theta^* \tau$. We now consider each feature in turn.
\subsection{Potentials with a ridge}
\label{sec:ridge}
The simplest possible ridge we can consider takes the form 
\be
W = W_0 + g \phi - \frac{1}{2} m_\chi^2 \chi^2 \,,
\ee
where $g$ and $m_\chi$ are taken to be positive and the $W_0$ term dominates. 
Since the potential is vacuum dominated and $\Omega \simeq 1$, 
the product-separable and sum-separable formulae for $\fnl$, $\tnl$ and $\gnl$ are identical,
because vacuum-domination also implies that $\ep \ll \eta_{ss}$.
To be consistent with our sign convention we stipulate that 
$\chi <0$ such that $W_{,i} >0$ and $0 \leq \theta \leq \pi/2$.
The initial conditions are fine-tuned such that the field initially moves almost parallel to 
the top of the ridge, with $\theta \gtrsim 0$. The isocurvature direction is therefore 
almost precisely the $\chi$ direction, and $\eta_{ss} \simeq -m_{\chi}^2/W_0$ is a constant.
This then gives
\be
\frac{6}{5}\fnl \simeq f \eta_\chi^* \,, \qquad
\tnl \simeq \tau {\eta_\chi^*}^2 \,, \qquad
\frac{54}{50} \gnl \simeq \frac{18}{5} \fnl \, \eta_\chi^* \,.
\label{eq:ridge_results1}
\ee
In this case $\gnl$ is subdominant, even to $\fnl$, and so we do not consider this further.
The peak value of $\fnl$ and $\tnl$ can then be found by maximising the 
functions $f$ and $\tau$ as $\theta$ varies. They peak when
$\theta^2 = \sqrt{3} \theta^*$ and  $\theta^2 = \sqrt{2} \theta^*$ respectively and so
\be
\frac{6}{5} \left. \fnl \right|_{\mathrm max} \simeq \frac{3 \sqrt{3}}{16} \frac{\eta_\chi^*}{\theta^*} \,, \qquad
 \left. \tnl\right|_{\mathrm max} \simeq \frac{4}{27} \frac{{\eta_\chi^*}^2}{\theta_*^2} \,.
\label{eq:ridge_results2}
\ee
The sign of $\fnl$ is negative due to the sign of $\eta_\chi^*$, and the
amplitude of $\fnl$ reaches its peak after $\tnl$.
The peak of $\tnl$ is only slightly larger than the square of
$\frac{6}{5} \left. \fnl \right|_{\mathrm max}$, but due to the difference in
peaking times it is quite possible to have $\tnl \gg |\fnl|$.

\subsection{Potentials with a valley}
\label{sec:valley}

The simplest possible vacuum-dominated valley takes the form 
\be
W = W_0 + \frac{1}{2} m_\phi^2 \phi^2 + \frac{1}{2} m_\chi^2 \chi^2 \,,
\ee
where $m_\phi$ and $m_\chi$ are taken to be positive and the $W_0$ term dominates. 
We again consider $\theta* \ll 1$, and assume that $m_\phi \gg m_\chi$, so that the 
initial motion is almost exactly parallel to the $\phi$ direction. Once again, therefore, 
the isocurvature direction is well approximated by the 
$\chi$ direction, and $\eta_{ss} = m_\chi^2/W_0$ is again constant.
This then gives expressions for $\fnl$, $\tnl$ and $\gnl$ which are identical to those in
Eq.~\eqref{eq:ridge_results1}, but we note that the signs of $\fnl$ and $\gnl$
are now reversed.

\subsection{Potentials with an inflection point}
\label{sec:inflectionpot}

The analysis of \S \ref{sec:gnltnl} with $\Omega \simeq 1$ illustrated how $\gnl$
is negligible for inflationary potentials that are approximately quadratic, but that
$\gnl \sim \tnl$ for potentials with cubic or higher order shapes. We now consider
the simplest such shape, with an inflection point of the form
\be
W = W_0 + g \phi + \frac{1}{6} \lambda \chi^3 \,.
\ee
The regime of interest is close to but below the inflection point with $|\chi| \ll 1$ such that
$\theta \ll 1$. The non-Gaussianity parameters are only large in the regimes where $f$ or
$\tau$ are large, which requires $|\chi| \gg |\chi^*|$. In this limit we find that the leading
order non-Gaussianity is given as
\be
\frac{6}{5}\fnl \simeq 2 f \eta_{ss} \,, \qquad
\tnl \simeq 4 \tau \eta_{ss}^2 \,, \qquad
\frac{54}{50} \gnl \simeq \frac{5}{8} \tnl \,.
\ee
For a general power-law potential, the numerical relation between $\gnl$ and $\tnl$ will
take a different value. We note that $\gnl$ simply follows the evolution of $\tnl$ and so
we need only calculate the peak values of $\fnl$ and $\tnl$. We find
\be
\frac{6}{5} \left. \fnl \right|_{\mathrm max} \simeq -1.08 \frac{|\xi_{sss}^*|}{{\theta^*}^{3/4}} \,, \qquad
 \left. \tnl\right|_{\mathrm max} \simeq 1.48  \frac{{\xi_{sss}^*}^2}{{\theta^*}^{3/2}} \,.
\ee
where these peaks respectively occur when $\theta^2 = \sqrt{3}\, \theta^*$ and
$\theta^2 = \sqrt{13/3}\, \theta^*$. 

\section{Concrete Models} \label{sec:models}

In the preceding discussion we developed a number of tools for
analysing a given two-field inflationary model. In this section we demonstrate the 
usefulness of these tools, by showing concrete examples of simple models
which can produce large values of $\tnl$ and/or $\gnl$.

\subsection{Two-field hybrid inflation} 
\label{sec:hybrid}
The original hybrid model was introduced by Linde~\cite{Linde:1991km} 
as an effectively single field model of inflation, where the job of ending inflation
was assigned to a second field, known as the waterfall field. The interesting
feature of this set up is that inflation does not end gracefully, 
but rather through a sudden transition, which allows more freedom
in constructing the potential that drives inflation. 
This model has been extended to the multi-field case, where two or more scalar 
fields drives inflation before the waterfall field ends it. In this section 
we consider a two-field hybrid model of inflation with the potential \cite{Lyth:2005fi}
\be
\label{eq:hybrid}
W =W_0 \left( 1 + \frac{1}{2} \eta_\phi \phi^2 + \frac{1}{2} \eta_\chi \chi^2 \right)
-\left(f\phi^2+g\phi\chi+h\chi^2\right)\frac{\psi^2}{2}+\frac{1}{2} m_{\psi}^2
\psi^2\, ,
\ee
where the waterfall field $\psi$ is held at $0$ during inflation, rolling
rapidly to it's true minimum when
\be
\label{eq:cond}
m_{\psi}^2 = f\phi^2+g\phi \chi+h\chi^2 \,.
\ee
Since inflation ends suddenly in this 
model before isocurvature modes decay, we need to account for the 
possibility that inflation might end on a surface other than one 
of constant energy density. This implies that the curvature perturbation 
produced by this model can be split into two parts,
\be
\label{eq:split}
 \zeta=\zeta_{\rm inf}+\zeta_{\rm e}\,,
\ee
where $\zeta_{\rm inf}$ is the part generated during inflation,
and $\zeta_e$ is the part generated from the end of inflation dynamics.
The condition \eqref{eq:cond} for the end of inflation also defines the surface on which inflation ends, 
which in this case is an ellipse. Varying $f$, $g$ and $h$ determines the
orientation and the eccentricity of the ellipse in the $\{\phi,\chi \}$ space; with $g=0$ 
defining an ellipse aligned with the axes, and $h=f$ with $g=0$ defining a circle. 
With this in mind, to determine which contribution in Eq. \eqref{eq:split} dominates
we look at the relationship between $f$, $h$ and $g$. For $g\gg f,h$ 
the $\zeta_{\rm e}$ term dominates over $\zeta_{\rm inf}$ \cite{Alabidi:2006wa,Salem:2005nd}.
We consider this case in the following subsection, and in the remainder of this section focus on
the scenario where $\zeta \simeq \zeta_{\rm inf}$ where the spectrum, bispectrum and trispectrum 
will evolve as described in our previous analysis for a two-field model, and there is no 
significant additional contribution  to $\zeta$ at the end of inflation. 
The characteristics and various observational predictions
of this model have been studied extensively in the literature, see for example
Refs. \cite{Lyth:2005qk,Alabidi:2006hg,Byrnes:2008wi,Byrnes:2008zy,Sasaki:2008uc, 
Naruko:2008sq,Huang:2009vk,Mulryne:2011ni,Elliston:2011et,Elliston:2011dr,
Abolhasani:2010kr,Suyama:2010uj,Mulryne:2009ci,Choi:2012he,Lyth:2012yp}.

We restrict our attention to the case in which $\eta_\phi$ and $\eta_\chi$ are both positive, 
and so the potential during inflation has a valley-like shape. This provides an illustrative example of 
our discussion in \S \ref{sec:valley}. 
If either $\eta_\phi$ or $\eta_\chi$ were negative, the potential would contain a ridge.
We further assume that $\eta_\phi > \eta_\chi$, and that the  
 initial conditions are such that $0 < \chi^* \ll \Mpl$, whilst
 $\phi^* \sim {\cal O} (\Mpl)$ is significantly displaced from zero. Hence, the trajectory 
 initially rolls towards the minimum along the $\phi$ axis before turning and rolling very slowly 
 towards the $\chi$ axis. 

These initial conditions imply a small $\theta^*$, and since the potential is
vacuum-dominated, $\alpha \simeq \theta$ and the heatmaps are particularly simple to interpret. 
The initial (horizon crossing) condition corresponds to a position on the diagonal in the bottom 
left hand corner of any given heatmap. As the trajectory begins its slow turn into the valley, 
the point on the heat-map proceeds vertically upwards and we can see that a strong 
non-Gaussian signal is expected during the early stages of the turn.
When the angle grows larger, the trajectory moves out of the regions 
of the heatmaps where $f$ and $\tau$ are large and so the non-Gaussianity decays. 
Since the large non-Gaussianity occurs during the early stages of the turn 
we see that $\eta_{ss} \approx \eta_\chi$ is a constant whilst all components 
of the $\xi_{ijk}^2$ parameter are zero.

\begin{figure}[t] 
    \center{\includegraphics[width = 8cm]{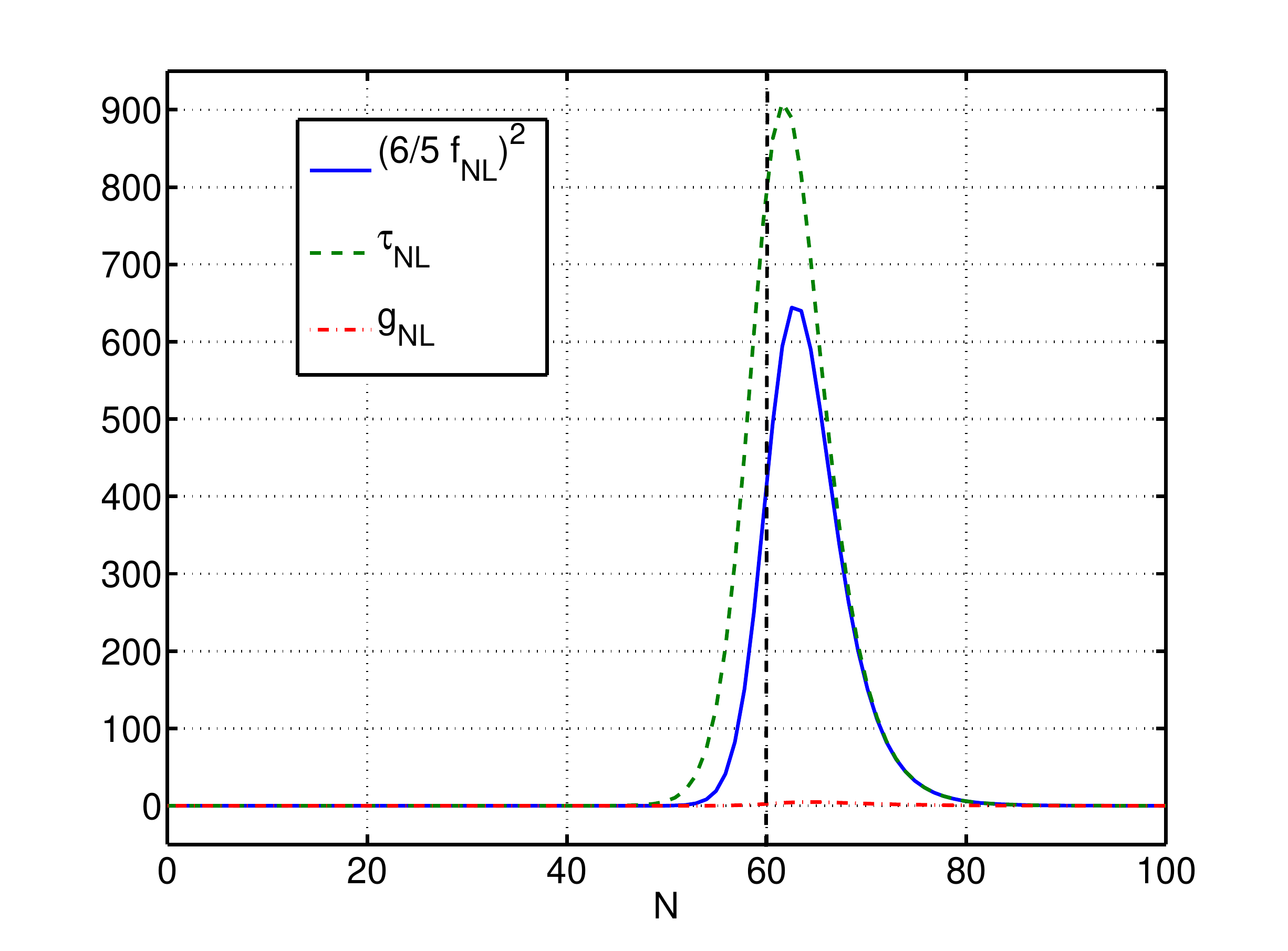}}
        \caption{Evolution of ${\rm sign} (\fnl) (6/5 \fnl)^2$ (solid blue line), 
$\tnl$ (dashed green line), and $\gnl$ (dot-dashed red line) for the two-field 
hybrid model described in the text, calculated using the analytic formulae presented in the appendix. 
The vertical dashed line at $60$ efolds indicates where we choose 
the waterfall field to terminate inflation. We also continue
the evolution past this point, as if the waterfall field were absent, to 
show the subsequent peak and decay of the non-Gaussianity.}
    \label{fig:2fHy}
       \end{figure}

In Fig.~\ref{fig:2fHy} we show the evolution of $\fnl$, $\tnl$ and $\gnl$ 
for the potential \eqref{eq:hybrid}. The parameter values used 
were $\eta_\phi=0.09$ and $\etc=0.0025$ with the initial conditions taken as $\phi^*=0.9 \Mpl$ and $\chi^*=0.001 \Mpl$. 
The evolution of the trispectrum is exactly as we expect: $\fnl$ and $\tnl$ both begin at 
negligible values, grow to large positive peaks and then decay again to negligible values.  
We confirm that the peak amplitudes are as expected
from Eq.~\eqref{eq:ridge_results2} in \S \ref{sec:valley} to within $10\%$. 
Furthermore, we see that $\tnl$ peaks before $\fnl$ in agreement with the discussion 
regarding Eq.~\eqref{eq:tnl} and also in agreement with the results in \S \ref{sec:valley}. 
This quadratic shape of the potential gives us a small $\gnl$ as discussed in \S \ref{sec:gnltnl}. 

After $60$ efolds, where we assume the evolution is terminated by the waterfall field, we find 
$n_s = 1.07$, which is clearly in violation
of the bounds on the spectral index placed by WMAP, which demands
a red spectral tilt for priors of zero running and tensor to scalar ratio. As this model
predicts neither a significant tensor to scalar ratio ($\gtrsim10^{-2}$) nor a large
negative running, we do not consider this model a viable candidate for inflation. 
Moreover, we know that there does not exist a choice
of parameters or initial conditions (where both fields are less than the Planck scale)
which can support a red tilt \cite{Alabidi:2006hg}. Nevertheless, we still consider the model 
a useful illustration of our analysis, and in particular of the a potential with a valley feature. 
It is also of note that a full treatment of the hybrid
model may lead to the production of cosmic strings (for example
see Refs.~\cite{Jeannerot:1997is, Jeannerot:2003qv}) and if cosmic strings were to contribute 
fractionally to the temperature power spectrum, then the best fit parameters would support a blue
spectral tilt (see for example Refs.~\cite{Bevis:2007gh,Bevis:2010gj}).

\subsubsection{The role of the end of inflation hypersurface}

As we have described, in addition to allowing for interesting 
phase space evolution during inflation, multi-field models 
of inflation can also have an end of inflation surface different from one of uniform energy density
\cite{Bernardeau:2002jf,Bernardeau:2004zz, Lyth:2005qk, Alabidi:2006wa,Salem:2005nd, Choi:2012he}, and this leads to the 
the extra contribution to the curvature perturbation $\zeta_{\rm e}$ discussed above. 
If this extra contribution dominates then the overall curvature perturbation can be 
significantly non-Gaussian and so the trispectrum may be enhanced.

Here we do not attempt a full analysis of the statistics when $\zeta_{\rm e}$ 
is accounted for, but rather offer one simple example. We assume $\zeta_{\rm e}\gg \zeta_{\rm inf}$, and 
consider the case where $\chi$ is held at zero during inflation
and the parameters $\eta_\phi$ and $\eta_\chi$ are equal.
The slow-roll parameters in this case are given as \cite{Alabidi:2006wa}
\be
\eta_{ss}=\eta_\chi=\eta_\phi\,, \qquad \ep=\frac{\eta_{ss}^2}{2}\phi^2\,,
\ee
and  $\gnl$ and $\tnl$ parameters at the end of inflation are found to be
\bea
\gnl &\simeq& \frac{50g}{54} \left(3\fnl\left(\eta_{ss} - 2\ep \right)+\eta_{ss}\right) \,, \nonumber \\
\tnl&\simeq&2\ep-\eta_{ss}+\frac{12\fnl}{5}\left(\frac{9\fnl}{10}-1\right)\,,\nonumber\\
\fnl&=&\frac{5f \eta_{ss}}{3g}\left(\frac{2h}{g}-\frac{g}{2f}\right) \,,
\eea
where all quantities are evaluated at the end of inflation.

We see that $\tnl$ closely follows the single-field result $\tnl\simeq (6\fnl/5)^2$
and so may be detectable if $\fnl$ is observably large as found in ref.~\cite{Alabidi:2006wa}.
When $\fnl \gg 1$, we see that $\gnl \propto g \fnl \times {\cal O}(\ep)$ and when $\fnl \ll 1$ then 
$\gnl \propto g \times {\cal O}(\ep)$. In both cases $\gnl$ is negligible unless one 
considers a very large coupling $g$ that is inconsistent with effective field theory
and so we see that, for this particular model, modifying the end of inflation hypersurface
produces non-Gaussianitites that are accounted for purely by the bispectrum.

In Ref.~\cite{Huang:2009vk}, the author relates the non-Gaussianity predictions from the end of
inflation hypersurface to the geometry of this surface. The author finds $\gnl$ may be
significant when the curvature changes rapidly along the surface.
Our findings here agree with this conclusion, because the parameter
choices $(g \gg 1)$ which lead to an enhanced $\gnl$ correspond to a large ellipticity
and have the trajectory lying very close to the semi-major axis.

\subsection{Axion Potential} 
The model with the potential
\be
\label{eq:2AxQuart}
W = \frac{1}{4} g \phi^4 + \Lambda^4 \left( 1 - \cos \left( 2 \pi \chi / f \right) \right)\,,
\ee
is inspired by the N-axion model studied in Refs. \cite{Kim:2010ud,Kim:2011je}. 
It is similar to the model introduced in Ref. \cite{Elliston:2011dr} 
as an example of a two-field potential which has an 
interesting transitory evolution as well as a large asymptotic value of 
non-Gaussianity at the adiabatic limit. Here we have chosen the $\phi$
field to self-interact through a quartic potential, which will allow a
choice of parameters and initial conditions for which $\zeta$ can become constant
during the slow-roll evolution, while still giving a value of the
spectral index consistent with observations\footnote{In Ref.~\cite{Elliston:2011et}, an
example was given in which a large non-Gaussianity was reached during slow-roll inflation.
We note, however, that the spectral index of this example was much too low to be consistent
with observation. By choosing a quartic potential in this work, our aim is to show that there is no
barrier in principal to having a model which reaches the adiabatic limit during slow-roll inflation
with a consistent value of $n_s$.}. 
The potential \eref{eq:2AxQuart} contains an infinite number of successive ridges and valleys. 
Taking initial conditions such that $\chi^*$ is sufficiently close to $f/2$, while $\phi$ starts its evolution 
at a large field value ($\phi^* \approx 23 \Mpl$ is required for roughly $60$ efolds), 
the trajectory is initially almost entirely in the $\phi$ direction and rolls along the ridge 
defined by $\chi = f/2$. The initial angle $\theta^*$ is therefore close to zero, and as the 
trajectory slowly turns off the ridge, $\theta$ increases.

From the the heatmaps, one can see that the initial growth of the angle $\theta$ will
lead to an increase in the magnitude of $\fnl$ and $\tnl$ 
as $\alpha$ passes up through the `hot-spot' region where $\fnl$ and $\tnl$ peak. 
Since the third derivative of the potential is insignificant in the early stages 
of this turn, $\gnl$ is not significant.
After the initial turn from the ridge-top, the trajectory will fall quickly down
the steep side of the ridge, before turning back towards its original direction 
as it reaches the valley bottom. This leads to a decrease of $\alpha$ back through the hot-spot region once more. 
Importantly, the potential $W$ drops by a small (but non-vanishing) amount between these two
turns. From Eq.\eqref{eq:alphadash2} this means that the initial growth of $\alpha$
is slightly larger than the subsequent decay and so $\alpha$ will not
make it to zero as $\theta \to 0$.
Instead, it will end up with a small positive value.
We therefore expect the non-Gaussianity
parameters to take large constant values asymptotically, where we also know that 
the phase space trajectory ends up in a valley and so the asymptotic formulae 
for $\gnl$ in Eq. \eqref{eq:gnl_tnl} will be an accurate prediction.

\begin{figure}[t] 
    \center{\includegraphics[width = 8cm]{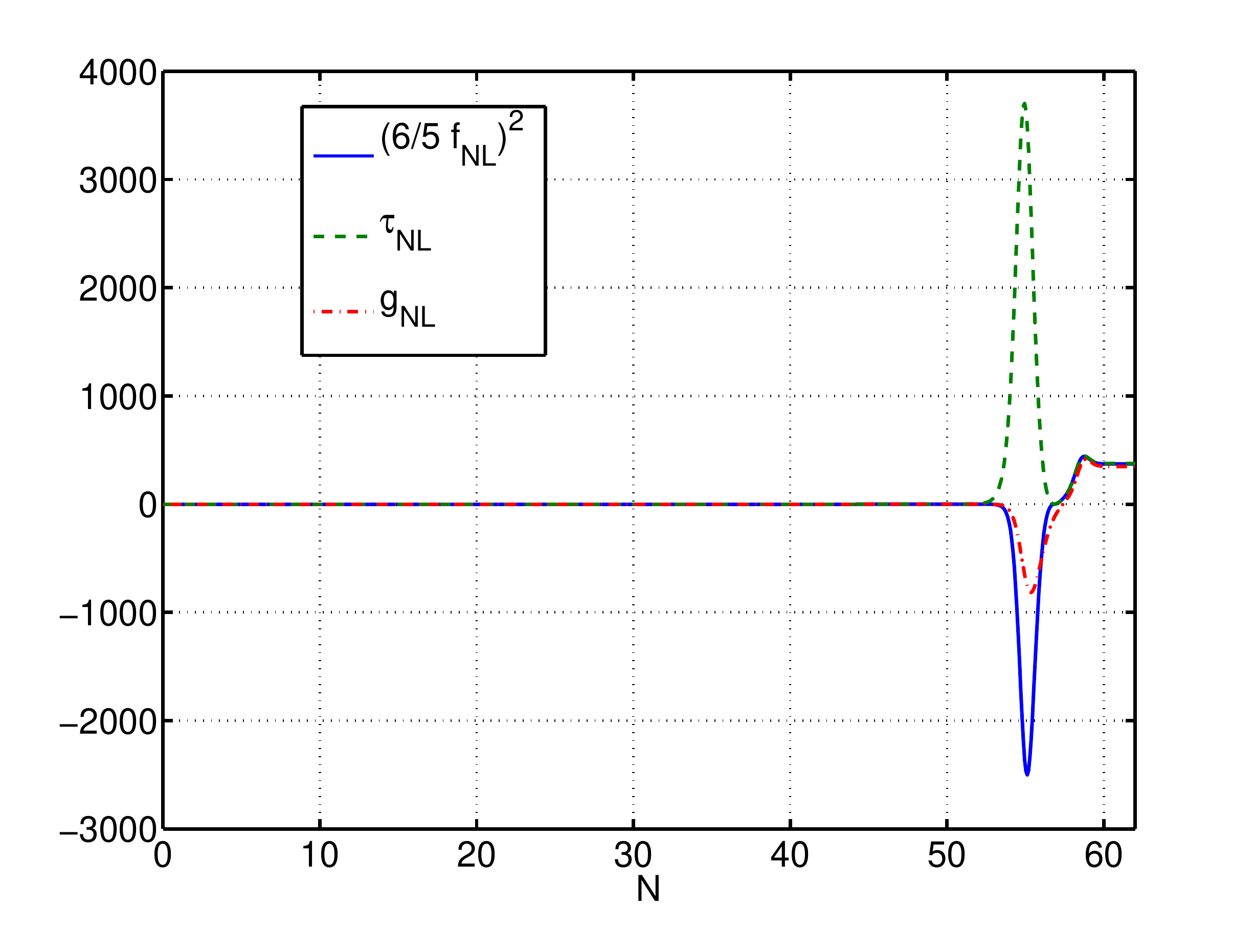}}
        \caption{Evolution of ${\rm sign} (\fnl) (6/5 \fnl)^2$ (solid blue line), $\tnl$ (dashed green line), and $\gnl$ (dot-dashed red line) for 
the axion model discussed in the text, calculated using the analytic formulae presented in the appendix. }
    \label{fig:axion}
       \end{figure}

In Fig.~\ref{fig:axion} we give the evolution of $\fnl$, $\tnl$ and $\gnl$ 
for a specific realisation of this case described above. 
The parameter values are $f=1\Mpl$ and $\Lambda^4/g = (25/2\pi)^2\Mpl^4$, with the overall
normalisation fixed to agree with WMAP7 power spectrum amplitude. 
The initial conditions are $\phi^*=22.5 \Mpl$ and $\chi^*=f/2-0.001\Mpl$
which gives us the HCA value of $\alpha$ as $0.022$ which is small in agreement with the above
discussion. In this case, the final constant spectral index has a value of
$n_s = 0.949$ which is within the WMAP7 $95\%$ contours.
The evolution is exactly as we expect, with $\fnl$ and $\tnl$ both beginning with
negligible values and $\fnl$ then evolving to a large negative peak while $\tnl$
grows to a large positive one, before both peaks decay. Despite the potential dropping quite 
significantly before the axion rolls, the peak values of
$\fnl$ and $\tnl$ are described by the formulae in \S \ref{sec:ridge} with about $30 \%$ accuracy.
The subsequent evolution is also interesting. As the 
trajectory evolves into the valley there is a positive spike in $\fnl$ and $\gnl$ typical of this 
evolution. Furthermore, $\gnl$ also increases significantly as we move away from 
the $\Omega=1$ regime and we see that there is a delay between the 
evolution of $\fnl$ and $\tnl$ as discussed in \S \ref{sec:ridge}. 
Finally, the constant asymptotic values of $\fnl$, $\tnl$ and $\gnl$ are
reached where we find Eq.~\eqref{eq:gnl_tnl} holds very well. 
In this case $\tnl \approx (6/5 \fnl)^2$, though this is not a necessary consequence of reaching the 
asymptotic regime, unlike the relation between $\tnl$ and $\gnl$.

\subsection{Inflection point model}\label{sec:inflection}

The discussion in \S \ref{sec:gnltnl} motivated sloping inflection features in the potential
as a scenario in which the condition $\gnl \gg \tnl$ might be found in an adiabatic regime 
where $\Omega \to 0$. We now present such a model, but note that 
the tuning required to find parameters such that $\gnl$ is both large and dominant in the
adiabatic limit is severe. We expect, therefore, that the simple relation \eqref{eq:gnl_tnl} will hold for
the vast majority of models.

We choose an inflection potential similar to that studied 
in Ref.~\cite{Elliston:2011et} which takes the form
\be           
W = W_0 \left( \frac{1}{4!} \phi^4 + V_0 + h \chi + \frac{1}{3!}\lambda \chi^3 +
\frac{1}{4!} \mu \chi^4\ \right)\,.
\ee                        
The value of $W_0$ is fixed by the WMAP power spectrum, and there is an
inflection point feature at $\chi = 0$. We further assume that $V_0$ and $\mu$ 
are fixed by the requirement that there is a minimum of the potential 
at $\chi_{\rm min} = -\rho$, where $\rho$ is taken to be positive. 
We note that beginning some way above/below the inflection point leads to a negative/positive 
asymptotic value of $\fnl$ as explored in Ref. \cite{Elliston:2011et}. 

From an exploration of the parameter space for this model we have found it very difficult to obtain 
a large and dominant $\gnl$ whilst satisfying the constraints of slow-roll.
We present a compromise scenario where the $\chi$ field rolls when $\ep \sim 0.3$. 
Such a model has parameter values
$h = 0.05 \Mpl^3$, $\lambda = 10^{4} \Mpl$ and $\rho=0.1 \Mpl$ and 
the evolution begins with $\phi^*=22.5 \Mpl$ and $\chi^*=0$~\footnote{ 
A realistic application of our formulae requires that the model is insensitive to
changes in the horizon crossing conditions within the range of the quantum scatter. 
Such changes do not affect the predictions of this model.}. 
We compare the evolution from our analytic expressions to the numerical non-slow-roll evolution
calculated using a finite-difference code identical to that used in Ref
\cite{Elliston:2011dr}.
In the non-slow-roll case, the scalar field velocity perturbations decouple from the 
field perturbations and so the phase space of the dynamics is enlarged.
We see the effect of this by the presence of oscillatory behaviour in the non-Gaussianity parameters
and in the spectral index. When the deviation from slow-roll is not too large, the non-slow-roll
evolution mimics a superposition of the the slow-roll evolution and these rapid oscillations. 
We therefore find that our analytic results are surprisingly applicable when slow-roll begins to break.
For this example, the analytic spectral index is $n_s=0.964$ and this is within $1\%$ of the
non-slow-roll value. Our calculation gives $\gnl=-432$ which agrees with the non-slow-roll code 
to within a factor of two. In general, if the evolution of the observables for a given model have not
settled down to a constant value before the end of slow-roll inflation then one must use numerical
methods to obtain an accurate result. However, a crude estimate is very easy to calculate 
through the slow-roll analysis. To conclude, in this final example we have tried to engineer a model 
to break from the simple $\gnl$--$\tnl$ result in Eq. \eqref{eq:gnl_tnl} but we have found that this is
very hard to achieve.

\section{Conclusions} \label{sec:conclusions}

In this work we have studied non-Gaussianities in two-field slow-roll inflation.
By extending and simplifying the `heatmap' analysis of Byrnes {\it et al.} 
\cite{Byrnes:2008wi} we have shown, for separable potentials, the regions of the 
parameter space that can give rise to large non-Gaussianities, both in the 
bi- and tri-spectra. This analysis also benefits from showing the explicit 
dependence of these non-Gaussianites on the shape of the inflationary potential, and 
the Horizon Crossing Approximation can be easily applied to our results.
The analyses in the paper allows us to make the following conclusions:

\begin{itemize}
\item We have found that the regions of the heatmap parameter space that produce large 
values of the bispectrum parameter $\fnl$ 
are also capable of producing large values of the trispectrum parameters $\tnl$ or $\gnl$. 
Our results confirm that a necessary requirement for a large 
local non-Gaussianity is that the horizon crossing field velocities must be dominated by one of the two fields.

\item The relationship between $\tnl$ and $\fnl$ is shown explicitly for separable potentials and is given in Eq.~\eqref{eq:tnl}.

\item We have found that in the adiabatic limit the non-Gaussianity parameters $\gnl$ and $\tnl$ 
can be related by Eq. \eqref{eq:gnl_tnl}, which is confirmed for a variety of models considered here.
This relation can be modified when the isocurvature third slow-roll parameter $\xi_{sss}^2$ 
is significant.

\item We have shown that the ridge and valley features in the inflationary potential can lead to large 
$\tnl$ and $\fnl$ parameters, with $\tnl$ being the first to peak.
$\gnl$ can be large if the potential is described by terms beyond quadratic order. It is possible, but
very hard to engineer, that $\gnl$ is the dominant statistic if inflation reaches a natural adiabatic limit.

\item We have employed the heatmaps in a dynamic way, and shown that they can be used
as qualitative tools, in order to understand and predict the evolution
of non-Gaussianities in different models of inflation.
\end{itemize}

Forthcoming Planck data will further constrain primordial 
non-Gaussianities. Since our analysis relates these observations to the shape
of the multi-field inflationary potential, it may be that if a signal is detected one could 
infer information about the 
inflationary potential from observation. This is an exciting possibility. Moreover, the non-Gaussianity
parameters considered here are not the only parameters that can act as signatures
for specific shapes in the potential. In particular, it would be valuable to extend 
our analysis to consider the running of the spectral index and the running of $\fnl$, $\gnl$ and $\tnl$, 
which we intend to return to in future work. 

\section{Acknowledgements} 

The authors would like to thank David Seery and Chris Byrnes for helpful discussions.
JE is supported by a Science and Technology Facilities Council studentship. LA is supported by the Japanese Society for the Promotion of Science (JSPS). 
IH, DJM and RT are supported by Science and Technology Facilities Council grant ST/J001546/1.

\appendix

\section{Summary of results for two-field $\delta N$ calculations.} \label{appendixA:fullcalcs}

\subsection{Sum-separable potentials}

This section follows from the work of Vernizzi and Wands \cite{Vernizzi:2006ve}
and considers a potential with the sum-separable form $W = U(\phi) + V(\chi)$. 
The first derivatives of $N$ are
\bea
\Mpl N_{,\phi}&=& \frac{u}{\sqrt{2 \epp^*}} \,, \qquad u = \frac{U^* + Z}{W^*}\,, \nonumber \\
\Mpl N_{,\chi}&=& \frac{v}{\sqrt{2 \epc^*}} \,, \qquad v = \frac{V^* - Z}{W^*}\,, \label{eq:ss_uv}
\eea
where
\be
\label{eq:Z}
Z = \frac{V \epp - U \epc}{\ep}\,.
\ee
For notational ease we have dropped the `$c$' label that is commonly attached to
quantities evaluated on a later-time uniform density hypersurface.
For the purposes of calculating the trispectrum we shall also need to decompose $\eta_{ij}$ and
$\xi_{ijk}^2$ into the kinematic basis on the uniform density hypersurface.
In the original $\{ \phi,\chi \}$ frame the potential is sum-separable and the only non-zero values of $\eta_{ij}$ and $\xi_{ijk}^2$ are those for which all of the indices are identical, 
which allows us to use the single-index notation $\etp$ and $\xi_\phi^2$.
After rotating into the kinematic basis with fields $\sigma$ and $s$, the potential generally loses it separable form and so it is necessary to use all of the indices.
The three $\eta$ components are
\bea
\eta_{\sigma \sigma} &=& \frac{\epp \etp + \epc \etc}{\ep}\,, \nonumber \\
\eta_{\sigma s} &=& \frac{\sqrt{\epp \epc}}{\ep} (\etc - \etp)\,, \nonumber \\
\eta_{ss} &=& \frac{\epc \etp + \epp \etc}{\ep}\,.
\eea
The components of $\xi_{ijk}^2$ in the kinematic frame are
\bea
\ep^{3/2}\, \xi^2_{\sigma \sigma \sigma} &=& \epc^{3/2}\, \xic^2 + \epp^{3/2} \, \xip^2 \,, \nonumber \\
\ep^{3/2}\, \xi^2_{\sigma \sigma s} &=& \epc \sqrt{\epp} \, \xic^2 - \epp \sqrt{\epc}\, \xip^2 \,, \nonumber \\
\ep^{3/2}\, \xi^2_{\sigma ss} &=& \epp \sqrt{\epc} \, \xic^2 + \epc \sqrt{\epp}\, \xip^2\,, \nonumber \\
\ep^{3/2}\, \xi^2_{sss} &=& \epp^{3/2} \, \xic^2 - \epc^{3/2}\, \xip^2 \,.
\eea
The second derivatives of $N$ follow by differentiation of Eqs. \eqref{eq:ss_uv}, giving 
\bea
\Mpl^2 N_{,\phi \phi}&=& 1 - \frac{u \etp^*}{2 \epp^*} + \frac{\A}{\epp^*}\,, \nonumber \\
\Mpl^2 N_{,\phi \chi}&=& - \frac{\A}{\sqrt{\epp^* \epc^*}}\,, \nonumber \\
\Mpl^2 N_{,\chi \chi}&=& 1 - \frac{v \etc^*}{2 \epc^*} + \frac{\A}{\epc^*}\,,
\eea
where
\be
\frac{\Mpl}{W^*}\sqrt{\frac{\epp^*}{2}} \frac{\partial Z}{\partial \phi^*} 
= - \frac{\Mpl}{W^*}\sqrt{\frac{\epc^*}{2}} \frac{\partial Z}{\partial \chi^*} 
= \A \equiv \frac{W^2}{W_*^2} \frac{\epp \epc}{\ep^2}
	\left( \eta_{ss} - \ep \right).
\label{eq:ss_A}
\ee
Taking the next derivative we find
\bea
\Mpl^3 N_{,\phi \phi \phi}&=& \frac{1}{\epp^* \sqrt{2 \epp^*}}
	\left(
	- \frac{u}{2} \sqrt{\frac{\epp^*}{\ep^*}} {\xip^*}^2  
	- \epp^* \etp^*
	+ u {\etp^*}^2
	- 3 \etp^* \A + \B^2 
	\right),\\
\Mpl^3 N_{,\phi \phi \chi}&=& \frac{1}{\epp^* \sqrt{2 \epc^*}}\left( \etp^* \A - \B^2 \right), \\
\Mpl^3 N_{,\phi \chi \chi}&=& \frac{1}{\epc^* \sqrt{2 \epp^*}}\left( \etc^* \A + \B^2 \right), \\
\Mpl^3 N_{,\chi \chi \chi}&=& \frac{1}{\epc^* \sqrt{2 \epc^*}}
	\left(
	- \frac{v}{2} \sqrt{\frac{\epc^*}{\ep^*}} {\xic^*}^2  
	- \epc^* \etc^*
	+ v {\etc^*}^2
	- 3 \etc^* \A - \B^2  
	\right),\\
\eea
where
\bea
\label{eq:ss_B}
\Mpl \sqrt{2 \epp^*} \frac{\partial \A}{\partial \phi_*} &\equiv& -4 \epp^* \A + \B^2 , \\
\B^2 &\equiv &  -\frac{W^3}{W_*^3} \frac{\sqrt{\epp \epc}^3}{\ep^3} \Big[
\xi_{sss}^2 + 2 \frac{\epp - \epc}{\sqrt{\epp \epc}} \eta_{ss} (\eta_{ss}-\ep)
- 2 \eta_{\sigma s} (\eta_{ss} + \ep)
\Big] . ~~~
\eea

The other derivative of $\A$ follows by a simple permutation of Eq.~\eqref{eq:ss_B} 
under the joint permutations $\{u \leftrightarrow v \}$ and $\{\phi \leftrightarrow \chi \}$
which has the effect of negating $\B^2$ whilst $\A$ remains unchanged. 
We take these results together to find
\bea
\label{eq:ss_params1}
n_s-1 = &- 4 \left(\frac{u^2}{\epp^*} + \frac{v^2}{\epc^*} \right)^{-1}
& \left[ 1 - \frac{u^2 \etp^*}{2 \epp^*} - \frac{v^2 \etc^*}{2 \epc^*} \right] - 2 \ep^* \nonumber \\
\fnl = & \frac{5}{6} \left(\frac{u^2}{\epp^*} + \frac{v^2}{\epc^*} \right)^{-2} 
&\left[ 
	2 \left( \frac{u^2}{\epp^*} +\frac{v^2}{\epc^*} \right)
	- \frac{u^3 \etp^*}{{\epp^*}^2 }
	- \frac{v^3 \etc^*}{{\epc^*}^2 } 
	+2 \left( \frac{u}{\epp^*} - \frac{v}{\epc^*} \right)^2 \A 
\right], \nonumber \\
\tnl =& 4 \left(\frac{u^2}{\epp^*} + \frac{v^2}{\epc^*} \right)^{-3} 
&\left[
	-\frac{u^3 \etp^*}{{\epp^*}^2}
	-\frac{v^3 \etc^*}{{\epc^*}^2}
	+\frac{u^4 {\etp^*}^2 }{4{\epp^*}^3 }
	+\frac{v^4 {\etc^*}^2 }{4{\epc^*}^3 } 
	+\frac{u^2}{\epp^*}
	+\frac{v^2}{\epc^*}\right. \nonumber \\
	&&\left. \quad - \frac{u^2}{ {\epp^*}^2} \left( \frac{u}{\epp^*} - \frac{v}{\epc^*} \right) \etp^* \A 
	- \frac{v^2}{ {\epc^*}^2} \left( \frac{v}{\epc^*} - \frac{u}{\epp^*} \right) \etc^* \A \right. \nonumber \\
	&& \left. \quad + 2 \left( \frac{u}{\epp^*} - \frac{v}{\epc^*} \right)^2 \A  
	+\left( \frac{u}{\epp^*} - \frac{v}{\epc^*}\right)^2 \left( \frac{1}{\epp^*} + \frac{1}{\epc^*}\right) \A^2
\right], \nonumber \\
\gnl =& \frac{50}{54}\left(\frac{u^2}{\epp^*} + \frac{v^2}{\epc^*} \right)^{-3} 
&\left[
	- \frac{u^3 \etp^*}{{\epp^*}^2}
	- \frac{v^3 \etc^*}{{\epc^*}^2}
	+ \frac{u^4 {\etp^*}^2}{{\epp^*}^3}
	+ \frac{v^4 {\etc^*}^2}{{\epc^*}^3}
	- \frac{1}{2}\frac{u^4 {\xip^*}^2}{{\epp^*}^2 \sqrt{\ep^* \, \epp^*} }
	- \frac{1}{2}\frac{v^4 {\xic^*}^2}{{\epc^*}^2 \sqrt{\ep^* \, \epc^*}}\right. \nonumber \\
	&&\left. \!\!\!\!\!\!\!\!\!\!\!\!\!\!\!\!\!\!\!\!\!\!\!\!\!\!\!\!\!\! -3 \frac{u^2}{{\epp^*}^2} \left(\frac{u}{\epp^*} - \frac{v}{\epc^*} \right) \etp^* \A 
	-3 \frac{v^2}{{\epc^*}^2} \left(\frac{v}{\epc^*} - \frac{u}{\epp^*} \right) \etc^* \A 
	+ \left(\frac{u}{\epp^*} - \frac{v}{\epc^*} \right)^3 \B^2.
\right] 
\eea

We now rewrite the horizon crossing slow-roll parameters in terms of their kinematic counterparts such as $\eta_{ss}^*$. 
There are three kinematic $\eta$ parameters and four $\xi^2$ parameters which means that there is no unique way in which to do this. 
We shall choose to rewrite $\etp^*$ and $\etc^*$ in terms of $\eta_{ss}^*$ 
and $\eta_{\sigma s}^*$ and ${\xip^*}^2$ and ${\xic^*}^2$ in terms 
of ${\xi_{sss}^*}^2$ and ${\xi_{\sigma ss}^*}^2$. This gives the relations
\bea
\etp = \eta_{ss} - \sqrt{\frac{\epp}{\epc}} \eta_{\sigma s} \, , \nonumber \\ 
\etc = \eta_{ss} + \sqrt{\frac{\epc}{\epp}} \eta_{\sigma s} \, , \nonumber \\
\xip^2 = \frac{\sqrt{\ep \epp}}{\epc}\xi_{\sigma ss}^2 - \sqrt{\frac{\ep}{\epc}} \xi_{sss}^2 \, , \nonumber \\
\xic^2 = \frac{\sqrt{\ep \epc}}{\epp}\xi_{\sigma ss}^2 + \sqrt{\frac{\ep}{\epp}} \xi_{sss}^2 \, .
\label{eq:kinematic_subs}
\eea
Substituting Eq.~\eqref{eq:kinematic_subs} into Eq.~\eqref{eq:ss_params1} and
simplifying we find
\begin{align}
n_s-1 &= 2(\eta_{ss}^* - \ep^*) - 2 \left(\frac{u^2}{\epp^*} +
\frac{v^2}{\epc^*} \right)^{-1} \left[ 2 + \frac{u -v}{\sqrt{\epp^* \epc^*}}
\eta_{\sigma s}^* \right]\,,  \nonumber \\
\fnl &= \frac{5}{6} \left(\frac{u^2}{\epp^*} + \frac{v^2}{\epc^*} \right)^{-2} 
\left[ 
	2 \left( \frac{u^2}{\epp^*} +\frac{v^2}{\epc^*} \right)
	- \left( \frac{u^3}{{\epp^*}^2} + \frac{v^3}{{\epc^*}^2} \right) \eta_{ss}^*  \right. \nonumber \\
	& \hspace{5cm} \quad\left. 
	+ \left( \frac{u^3}{\epp^*} - \frac{v^3}{\epc^*} \right) \frac{\eta_{\sigma s}^*}{\sqrt{\epp^* \epc^*}}
	+2 \left( \frac{u}{\epp^*} - \frac{v}{\epc^*} \right)^2 \A 
\right], \nonumber \displaybreak[0]\\
\tnl &= \left(\frac{u^2}{\epp^*} + \frac{v^2}{\epc^*} \right)^{-3} 
\left[
	\left( \frac{u^4}{{\epp^*}^3} + \frac{v^4}{{\epc^*}^3} \right){\eta_{ss}^*}^2
	-2\left( \frac{u^4}{{\epp^*}^2} - \frac{v^4}{{\epc^*}^2} \right)\frac{\eta_{ss}^* \eta_{\sigma s}^*}{\sqrt{\epp^* \epc^*}} \right. \nonumber \\
	& \left.
	+\left( \frac{u^4}{\epp^*} + \frac{v^4}{\epc^*} \right)\frac{{\eta_{\sigma s}^*}^2}{\epp^* \epc^*}
	-4 \left(\frac{u^3}{{\epp^*}^2} + \frac{v^3}{{\epc^*}^2} \right)\eta_{ss}^* 
	+ 4\left(\frac{u^3}{\epp^*} - \frac{v^3}{\epc^*} \right)\frac{\eta_{\sigma s}^*}{\sqrt{\epp^* \epc^*}} \right. \nonumber \\
	&\left. 	
	+4\left(\frac{u^2}{\epp^*} + \frac{v^2}{\epc^*}\right) 
	- 4\left(\frac{u}{\epp^*}-\frac{v}{\epc^*}\right)^2 \left( \frac{u}{\epp^*} + \frac{v}{\epc^*} \right) \eta_{ss}^* \A 
	+ 4\left(\frac{u}{\epp^*}-\frac{v}{\epc^*}\right) \left( \frac{u^2}{\epp^*} + \frac{v^2}{\epc^*} \right) \frac{\eta_{\sigma s}^* \A}{\sqrt{\epp^* \epc^*}} \right. \nonumber \\
	& \left. \hspace{4cm} + 8 \left( \frac{u}{\epp^*} - \frac{v}{\epc^*}
\right)^2 \A  
	+4\left( \frac{u}{\epp^*} - \frac{v}{\epc^*}\right)^2 \frac{\ep^* \A^2}{\epp^* \epc^*}
\right], \nonumber \displaybreak[0]\\
\gnl &= \frac{50}{54}\left(\frac{u^2}{\epp^*} + \frac{v^2}{\epc^*} \right)^{-3} 
\left[
	\left( \frac{u^4}{{\epp^*}^3} + \frac{v^4}{{\epc^*}^3} \right){\eta_{ss}^*}^2
	-2\left( \frac{u^4}{{\epp^*}^2} - \frac{v^4}{{\epc^*}^2} \right)\frac{\eta_{ss}^* \eta_{\sigma s}^*}{\sqrt{\epp^* \epc^*}} \right. \nonumber \\
	& \left.
	+\left( \frac{u^4}{\epp^*} + \frac{v^4}{\epc^*} \right)\frac{{\eta_{\sigma s}^*}^2}{\epp^* \epc^*}
	- \left(\frac{u^3}{{\epp^*}^2} + \frac{v^3}{{\epc^*}^2} \right)\eta_{ss}^* 
	+\left(\frac{u^3}{\epp^*} - \frac{v^3}{\epc^*} \right)\frac{\eta_{\sigma
s}^*}{\sqrt{\epp^* \epc^*}} \right. \nonumber \displaybreak[0]\\
	&\left. 	
	+\frac{1}{2} \left(\frac{u^4}{{\epp^*}^2} - \frac{v^4}{{\epc^*}^2} \right)\frac{{\xi_{sss}^*}^2}{\sqrt{\epp^* \epc^*}}
	-\frac{1}{2} \left(\frac{u^4}{\epp^*} + \frac{v^4}{\epc^*} \right)\frac{{\xi_{\sigma ss}^*}^2}{\epp^* \epc^*}
	\right. \nonumber \\
	&\left. 	
	- 3\left(\frac{u}{\epp^*}-\frac{v}{\epc^*}\right)^2 \left( \frac{u}{\epp^*} + \frac{v}{\epc^*} \right) \eta_{ss}^* \A 
	+ 3\left(\frac{u}{\epp^*}-\frac{v}{\epc^*}\right) \left( \frac{u^2}{\epp^*} + \frac{v^2}{\epc^*} \right) \frac{\eta_{\sigma s}^* \A}{\sqrt{\epp^* \epc^*}}  + \left( \frac{u}{\epp^*} - \frac{v}{\epc^*} \right)^3 \B^2  
\right] \,. \nonumber \displaybreak[0]\\
\label{eq:ss_params}
\end{align}

\subsection{Product-separable potentials}

For a potential with the two-field product-separable form $W = U(\phi)V(\chi)$, 
the first derivatives of $N$ are
\bea
\Mpl N_{,\phi}&=& \frac{u}{\sqrt{2 \epp^*}}\,, \qquad u = \frac{\epp}{\ep}\,, \nonumber \\
\Mpl N_{,\chi}&=& \frac{v}{\sqrt{2 \epc^*}} \,, \qquad v = \frac{\epc}{\ep} \label{eq:ps_uv}\,.
\eea
In the $\{ \phi,\chi \}$ frame the potential is product-separable which means that any of the parameters $\eta_{ij}$ or $\xi_{ijk}^2$ 
with mixed derivatives can be written in terms of lower-order slow-roll parameters that do not have mixed derivatives.
This again allows us to use the single-index notation $\etp$ and $\xi_\phi^2$ for the remaining terms.
In the kinematic basis the three $\eta$ components are
\bea
\eta_{\sigma \sigma} &=& \frac{\epp \etp + \epc \etc + 4 \epp \epc}{\ep} \,, \nonumber \\
\eta_{\sigma s} &=& \frac{\sqrt{\epp \epc}}{\ep} \left[(\etc -2\epc) - (\etp - 2\epp) \right]\,, \nonumber \\
\eta_{ss} &=& \frac{\epc \etp + \epp \etc - 4 \epp \epc}{\ep} \,.
\eea
The components of the $\xi_{ijk}^2$ tensor are
\bea
\ep^{3/2}\, \xi^2_{\sigma \sigma \sigma} &=& \epc^{3/2}\, \xic^2 + \epp^{3/2} \, \xip^2 +6\epp \epc \sqrt{\ep} (\etp + \etc)\,, \nonumber \\
\ep^{3/2}\, \xi^2_{\sigma \sigma s} &=& \epc \sqrt{\epp} \, \xic^2 - \epp \sqrt{\epc}\, \xip^2 + 2 \sqrt{\ep \epp \epc} \big[ (\epp-2\epc)\etp-(\epc-2\epp)\etc \big]\,, \nonumber \\
\ep^{3/2}\, \xi^2_{\sigma ss} &=& \epp \sqrt{\epc} \, \xic^2 + \epc \sqrt{\epp}\, \xip^2 + 2 \sqrt{\ep} \big[ (\epp-2\epc)\epp \etc+(\epc-2\epp) \epc \etp \big] \,, \nonumber \\
\ep^{3/2}\, \xi^2_{sss} &=& \epp^{3/2} \, \xic^2 - \epc^{3/2}\, \xip^2 +6 \sqrt{\ep \epp \epc} (\epc \etp - \epp \etc) \,.
\eea
The second derivatives of $N$ follow by differentiation of Eqs. \eqref{eq:ps_uv}, giving
\bea
\Mpl^2 N_{,\phi \phi}&=& u - \frac{u \etp^*}{2 \epp^*} + \frac{\A_P}{\epp^*}, \nonumber \\
\Mpl^2 N_{,\phi \chi}&=& - \frac{\A_P}{\sqrt{\epp^* \epc^*}}, \nonumber \\
\Mpl^2 N_{,\chi \chi}&=& v - \frac{v \etc^*}{2 \epc^*}  + \frac{\A_P}{\epc^*},
\eea
where we have substituted
\be
\A_P \equiv u v \, \eta_{ss}\,,
\label{eq:ps_A}
\ee
to put these equations into a form similar to that found for the sum-separable potential.
Taking the next derivative we find
\bea
\Mpl^3 N_{,\phi \phi \phi}&=& \frac{1}{\epp^* \sqrt{2 \epp^*}}
	\left(
	- \frac{u}{2} \sqrt{\frac{\epp^*}{\ep^*}} {\xip^*}^2  
	- u \epp^* \etp^*
	+ u {\etp^*}^2
	- 3 (\etp^* - 2 \epp^*) \A_P 
	+ \B_P^2
	\right),\\
\Mpl^3 N_{,\phi \phi \chi}&=& \frac{1}{\epp^* \sqrt{2 \epc^*}}\Big( (\etp^* - 2 \epp^*) \A_P - \B_P^2 \Big), \\
\Mpl^3 N_{,\phi \chi \chi}&=& \frac{1}{\epc^* \sqrt{2 \epp^*}}\Big( (\etc^* - 2 \epc^*) \A_P + \B_P^2 \Big), \\
\Mpl^3 N_{,\chi \chi \chi}&=& \frac{1}{\epc^* \sqrt{2 \epc^*}}
	\left(
	- \frac{v}{2} \sqrt{\frac{\epc^*}{\ep^*}} {\xic^*}^2  
	- v \epc^* \etc^*
	+ v {\etc^*}^2
	- 3 (\etc^* - 2\epc^*) \A_P 
	- \B_P^2
	\right),\\
\eea
where
\bea
\Mpl \sqrt{2 \epp^*} \frac{\partial \A_P}{\partial \phi_*} &\equiv &
- \Mpl \sqrt{2 \epc^*} \frac{\partial \A_P}{\partial \chi_*}
\equiv \B_P^2 \,, \nonumber \\
\B_P^2 &\equiv& -\sqrt{uv}^3 \Big[
\xi_{sss}^2 +2 \frac{\epp-\epc}{\sqrt{\epp \epc}} \eta_{ss}^2 - 2 \eta_{\sigma s} \eta_{ss}
\Big]\,.
\label{eq:ps_B}
\eea
As with the sum-separable case, $\A_P$ is symmetric under switching the fields whilst $\B_P^2$ is anti-symmetric. 
We take these results together to find
\bea
\label{eq:ps_params1}
n_s-1 = &- 4 \left(\frac{u^2}{\epp^*} + \frac{v^2}{\epc^*} \right)^{-1}
& \left[ 1 - 2 uv - \frac{u^2 \etp^*}{2 \epp^*} - \frac{v^2 \etc^*}{2 \epc^*} \right] - 2 \ep^* \nonumber \\
\fnl = & \frac{5}{6} \left(\frac{u^2}{\epp^*} + \frac{v^2}{\epc^*} \right)^{-2} 
&\left[ 
	2 \left( \frac{u^3}{\epp^*} +  \frac{v^3}{\epc^*} \right)
	- \frac{u^3 \etp^*}{{\epp^*}^2 }
	- \frac{v^3 \etc^*}{{\epc^*}^2 } 
	+2 \left( \frac{u}{\epp^*} - \frac{v}{\epc^*} \right)^2 \A_P 
\right], \nonumber \\
\tnl =& 4 \left(\frac{u^2}{\epp^*} + \frac{v^2}{\epc^*} \right)^{-3} 
&\left[
	-\frac{u^4 \etp^*}{{\epp^*}^2}
	-\frac{v^4 \etc^*}{{\epc^*}^2}
	+\frac{u^4 {\etp^*}^2 }{4{\epp^*}^3 }
	+\frac{v^4 {\etc^*}^2 }{4{\epc^*}^3 } 
	+\frac{u^4}{\epp^*}
	+\frac{v^4}{\epc^*}\right. \nonumber \\
	&&\left. \quad - \frac{u^2}{ {\epp^*}^2} \left( \frac{u}{\epp^*} - \frac{v}{\epc^*} \right) \etp^* \A_P
	- \frac{v^2}{ {\epc^*}^2} \left( \frac{v}{\epc^*} - \frac{u}{\epp^*} \right) \etc^* \A_P \right. \nonumber \\
	&& \left. \quad + 2 \left( \frac{u^2}{\epp^*} - \frac{v^2}{\epc^*} \right) \left( \frac{u}{\epp^*} - \frac{v}{\epc^*} \right) \A_P  
	+\left( \frac{u}{\epp^*} - \frac{v}{\epc^*}\right)^2 \left( \frac{1}{\epp^*} + \frac{1}{\epc^*}\right) \A_P^2
\right], \nonumber \\
\gnl =& \frac{50}{54}\left(\frac{u^2}{\epp^*} + \frac{v^2}{\epc^*} \right)^{-3} 
&\left[
	- \frac{u^4 \etp^*}{{\epp^*}^2}
	- \frac{v^4 \etc^*}{{\epc^*}^2}
	+ \frac{u^4 {\etp^*}^2}{{\epp^*}^3}
	+ \frac{v^4 {\etc^*}^2}{{\epc^*}^3}
	- \frac{1}{2}\frac{u^4 {\xip^*}^2}{{\epp^*}^2 \sqrt{\ep^* \, \epp^*} }
	- \frac{1}{2}\frac{v^4 {\xic^*}^2}{{\epc^*}^2 \sqrt{\ep^* \, \epc^*}}\right. \nonumber \\
	&&\left. \quad -3 \frac{u^2}{{\epp^*}^2} \left(\frac{u}{\epp^*} - \frac{v}{\epc^*} \right) \etp^* \A_P 
	-3 \frac{v^2}{{\epc^*}^2} \left(\frac{v}{\epc^*} - \frac{u}{\epp^*} \right) \etc^* \A_P \right. \nonumber \\
	&&\left. \quad +6 \left( \frac{u^2}{\epp^*} - \frac{v^2}{\epc^*}\right) \left( \frac{u}{\epp^*} - \frac{v}{\epc^*}\right) \A_P
	+ \left(\frac{u}{\epp^*} - \frac{v}{\epc^*} \right)^3 \B_P^2.
\right] .
\eea

One can rewrite the horizon crossing slow-roll parameters in terms of their kinematic counterparts as
\bea
\etp &=& \eta_{ss} + 2 \epp - \sqrt{\frac{\epp}{\epc}} \eta_{\sigma s} \, , \nonumber \\ 
\etc &=& \eta_{ss} + 2 \epc + \sqrt{\frac{\epc}{\epp}} \eta_{\sigma s} \, , \nonumber \\
\xip^2 &=& \frac{\sqrt{\ep \epp}}{\epc}\xi_{\sigma ss}^2 - \sqrt{\frac{\ep}{\epc}} \xi_{sss}^2 
-2 \frac{\sqrt{\ep \epp}}{\epc}(\epp \etc - 2 \epc \etp)\, , \nonumber \\
\xic^2 &=& \frac{\sqrt{\ep \epc}}{\epp}\xi_{\sigma ss}^2 + \sqrt{\frac{\ep}{\epp}} \xi_{sss}^2 
-2 \frac{\sqrt{\ep \epc}}{\epp}(\epc \etp - 2 \epp \etc)\, .
\label{eq:kinematic_subs_2}
\eea
Substituting Eq.~\eqref{eq:kinematic_subs_2} into Eq.~\eqref{eq:ps_params1} and
simplifying we find
\bea
n_s-1 &=& 2(\eta_{ss}^* - \ep^*) - 2 \left(\frac{u^2}{\epp^*} + \frac{v^2}{\epc^*} \right)^{-1} \frac{u -v}{\sqrt{\epp^* \epc^*}} \eta_{\sigma s}^* \,,  \nonumber \\
\fnl &=& \frac{5}{6} \left(\frac{u^2}{\epp^*} + \frac{v^2}{\epc^*} \right)^{-2} 
\left[ 
	- \left( \frac{u^3}{{\epp^*}^2} + \frac{v^3}{{\epc^*}^2} \right) \eta_{ss}^*  	+ \left( \frac{u^3}{\epp^*} - \frac{v^3}{\epc^*} \right) \frac{\eta_{\sigma s}^*}{\sqrt{\epp^* \epc^*}}
	+2 \left( \frac{u}{\epp^*} - \frac{v}{\epc^*} \right)^2 \A_P
\right], \nonumber \\
\tnl &=& \left(\frac{u^2}{\epp^*} + \frac{v^2}{\epc^*} \right)^{-3} 
\left[
	\left( \frac{u^4}{{\epp^*}^3} + \frac{v^4}{{\epc^*}^3} \right){\eta_{ss}^*}^2
	-2\left( \frac{u^4}{{\epp^*}^2} - \frac{v^4}{{\epc^*}^2} \right)\frac{\eta_{ss}^* \eta_{\sigma s}^*}{\sqrt{\epp^* \epc^*}} \right. \nonumber \\
	&& \left. \hspace{2.5cm}
	+\left( \frac{u^4}{\epp^*} + \frac{v^4}{\epc^*} \right)\frac{{\eta_{\sigma s}^*}^2}{\epp^* \epc^*}
	- 4\left(\frac{u}{\epp^*}-\frac{v}{\epc^*}\right)^2 \left( \frac{u}{\epp^*} + \frac{v}{\epc^*} \right) \eta_{ss}^* \A_P 	
	\right. \nonumber \\
	&&\left. \hspace{2.5cm}	
	+ 4\left(\frac{u}{\epp^*}-\frac{v}{\epc^*}\right) \left( \frac{u^2}{\epp^*} + \frac{v^2}{\epc^*} \right) \frac{\eta_{\sigma s}^* \A_P}{\sqrt{\epp^* \epc^*}} 
	+4\left( \frac{u}{\epp^*} - \frac{v}{\epc^*}\right)^2 \frac{\ep^* \A_P^2}{\epp^* \epc^*}
\right], \nonumber \\
\gnl &=& \frac{50}{54}\left(\frac{u^2}{\epp^*} + \frac{v^2}{\epc^*} \right)^{-3} 
\left[
	\left( \frac{u^4}{{\epp^*}^3} + \frac{v^4}{{\epc^*}^3} \right){\eta_{ss}^*}^2
	-2 \left( \frac{u^4}{{\epp^*}^2} - \frac{v^4}{{\epc^*}^2} \right) \frac{\eta_{ss}^* \eta_{\sigma s}^*}{\sqrt{\epp^* \epc^*}} \right. \nonumber \\
	&& \left.
	+\left( \frac{u^4}{\epp^*} + \frac{v^4}{\epc^*} \right)\frac{{\eta_{\sigma s}^*}^2}{\epp^* \epc^*}
	+ \left(\frac{u^4}{\epp^*} + \frac{v^4}{\epc^*} \right)\frac{\ep^* \eta_{ss}^*}{\epp^* \epc^*}  \right. \nonumber \\
	&&\left. 	
	+\frac{1}{2} \left(\frac{u^4}{{\epp^*}^2} - \frac{v^4}{{\epc^*}^2} \right) \frac{{\xi_{sss}^*}^2}{\sqrt{\epp^* \epc^*}}
	-\frac{1}{2} \left(\frac{u^4}{\epp^*} + \frac{v^4}{\epc^*} \right) \frac{{\xi_{\sigma ss}^*}^2}{ \epp^* \epc^*}
	\right. \nonumber \\
	&&\left. 	
	- 3\left(\frac{u}{\epp^*}-\frac{v}{\epc^*}\right)^2 \left( \frac{u}{\epp^*} + \frac{v}{\epc^*} \right) \eta_{ss}^* \A_P 
	+ 3\left(\frac{u}{\epp^*}-\frac{v}{\epc^*}\right) \left( \frac{u^2}{\epp^*} + \frac{v^2}{\epc^*} \right) \frac{\eta_{\sigma s}^* \A_P}{\sqrt{\epp^* \epc^*}}  + \left( \frac{u}{\epp^*} - \frac{v}{\epc^*} \right)^3 \B_P^2  
\right] \,. \nonumber \\
\label{eq:ps_params}
\eea

\section{Simplification of Trispectrum expressions}
\label{appendixB:fullcalcs}

The function $|\tau_6|$, as defined in Eq.~\eqref{eq:ss_tnl_para}, is bounded well within our
limit of ten and so this term may be immediately neglected.
Furthermore, we can use Eq.~\eqref{eq:eta_relationship} to manipulate various
other terms.
In the sum-separable case we find
\bea
\tau_2 \eta_{ss}^* \eta_{\sigma s}^* &=& \Big[\sin 2 \theta^*  \tau_2\Big] \eta_{ss}^* \eta_{\sigma s}^* 
+ \Big[(1-\sin 2\theta^*) \tau_2 \frac{1}{2} \tan 2 \theta^* \Big] \eta_{ss}^* (\eta_{ss}^* -\eta_{\sigma \sigma}^*)\,, \nonumber \\
\tau_3 {\eta_{\sigma s}^*}^2 &=& \Big[\sin^2 2\theta^* \tau_3\Big] {\eta_{\sigma s}^*}^2 
+ \Big[(1-\sin^2 2\theta^*) \tau_3 \frac{1}{4} \tan^2 2 \theta^* \Big]  (\eta_{ss}^* -\eta_{\sigma \sigma}^*)^2 \,, \nonumber \\
\tau_5 \ep^* \eta_{\sigma s}^* &=&  \Big[\sin 2 \theta^* \tau_5 \Big] \ep^* \eta_{\sigma s}^* 
+\Big[(1-\sin 2 \theta^*) \tau_5 \frac{1}{2} \tan 2\theta^* \Big] \ep^* (\eta_{ss}^*-\eta_{\sigma \sigma}^*)\,, \nonumber \\
- \tau_8 \, \Omega \, \eta_{\sigma s}^* (\eta_{ss}-\ep) &=&
-\Big[\sin^2 2 \theta^* \tau_8\Big] \, \Omega \, \eta_{\sigma s}^* (\eta_{ss}-\ep) \nonumber \\
&& \quad
-\Big[(1-\sin 2 \theta^*) \tau_8 \frac{1}{2} \tan 2\theta^* \Big] \, \Omega \, (\eta_{ss}^*-\eta_{\sigma \sigma}^*) (\eta_{ss}-\ep) \,,
\eea
and we find that all of the functions in square brackets never have a magnitude greater than ten
and so these terms represent variations in the trispectrum 
that are significantly smaller than observables will ever probe. 
One can easily check that these results follow analogously for product-separable potentials and so these terms may be neglected for both types of separable potential.

Analogously to Eq.~\eqref{eq:eta_relationship} there exist formulae relating the
various $\xi_{ijk}^2$ components. For sum-separable potentials we have
\bea
\xi_{\sigma ss}^2 &=& \frac{1}{2} \tan 2 \theta (\xi_{sss}^2-\xi_{\sigma \sigma s}^2)\,, \label{eq:ss_xi_relation1} \\
\xi_{\sigma \sigma s}^2 &=& \frac{1}{2} \tan 2 \theta (\xi_{\sigma ss}^2-\xi_{\sigma \sigma \sigma}^2)\,,
\label{eq:ss_xi_relation2}
\eea
whereas for product-separable potentials these relations are of the form
\bea
\xi_{\sigma ss}^2 &=& \frac{1}{2} \tan 2 \theta \, (\xi_{sss}^2-\xi_{\sigma \sigma s}^2) + 2 \ep \eta_{ss} + 2 \tan 2 \theta \, \ep \, \eta_{\sigma s}\,,  \label{eq:ps_xi_relation1} \\
\xi_{\sigma \sigma s}^2 &=& \frac{1}{2} \tan 2 \theta \, (\xi_{\sigma ss}^2-\xi_{\sigma \sigma \sigma}^2) - 2 \ep \, \eta_{\sigma s} 
+ 2 \tan 2 \theta \, \ep \, (\eta_{ss}+\ep)\,. \label{eq:ps_xi_relation2}
\eea
Using Eqs.~\eqref{eq:ss_xi_relation1} and \eqref{eq:ss_xi_relation2}
we can then simplify the full expression for $\gnl$ in
Eq.~\eqref{eq:ss_gnl_full} as
\bea
\frac{1}{4} \tau_2 {\xi_{sss}^*}^2 -\frac{1}{2} \tau_3 {\xi_{\sigma ss}^*}^2 
&=& \frac{1}{4} \tau_2 {\xi_{sss}^*}^2 - \sin^2 2 \theta^* \frac{1}{2} \tau_3 {\xi_{\sigma ss}^*}^2 
-(1-\sin^2 2 \theta^*) \frac{1}{2} \tau_3 \nonumber \\
&& \quad \times \left(\frac{1}{2} \tan 2 \theta^* {\xi_{sss}^*}^2 
- \frac{1}{4} \tan^2 2 \theta^* \left({\xi_{\sigma ss}^*}^2 
- {\xi_{\sigma \sigma \sigma}^*}^2\right) \right) \nonumber \\
&=& -g_4{\xi_{sss}^*}^2 - \Big[ \frac{1}{2} \tau_3 \sin^2 2 \theta^* \Big]
\left(\frac{3}{4}{\xi_{\sigma ss}^*}^2 + \frac{1}{4}{\xi_{\sigma \sigma \sigma}^*}^2  \right) \nonumber \\
&\simeq & -g_4{\xi_{sss}^*}^2 \,,
\eea
where we have defined $g_4 = \frac{1}{4} \left(\tau_3 \sin 2 \theta^* \cos 2 \theta^* -\tau_2 \right)$ in the expressions above. 
The term in square brackets in the last line can never be large and so is neglected.

The product-separable case follows similarly, yielding the same answer, however
the calculation is unsurprisingly more involved. Manipulating three of the
terms in Eq.~\eqref{eq:ps_gnl_full} for $\gnl$ by using Eq.
\eqref{eq:ps_xi_relation1} we find
\bea
\frac{1}{4} \tau_2 {\xi_{sss}^*}^2 + \tau_3 \ep^* \eta_{ss}^* -\frac{1}{2} \tau_3 {\xi_{\sigma ss}^*}^2 
&=& -g_4{\xi_{sss}^*}^2 - \Big[ \frac{1}{2} \tau_3 \sin^2 2 \theta^* \Big]
\left( {\xi_{\sigma ss}^*}^2 - 2 \ep^* \eta_{ss}^*\right) \nonumber \\
&& \quad + \frac{1}{4} \tau_3 \sin 2 \theta^* \cos 2 \theta^* \left({\xi_{\sigma \sigma s}^*}^2 - 4 \ep^* \eta_{\sigma s}^* \right)\,. \quad \quad \quad
\label{eq:ps_xi_working}
\eea
We now expand the last term of Eq.~\eqref{eq:ps_xi_working} by substituting for
${\xi_{\sigma \sigma s}^*}^2$ using Eq.~\eqref{eq:ps_xi_relation2}. We also 
use Eq.~\eqref{eq:eta_relationship} to rewrite $\eta_{\sigma s}^*$ as $\sin 2
\theta^* \eta_{\sigma s}^* + (1-\sin 2 \theta^*) \frac{1}{2} \tan 2
\theta^*(\eta_{ss} - \eta_{\sigma \sigma}+2\ep)$ and so we ultimately find
\bea
\frac{1}{4} \tau_2 {\xi_{sss}^*}^2 + \tau_3 \ep^* \eta_{ss}^* -\frac{1}{2} \tau_3 {\xi_{\sigma ss}^*}^2 
&=&-g_4{\xi_{sss}^*}^2 
- \Big[ \frac{1}{2} \tau_3 \sin^2 2 \theta^* \Big]
\left( \frac{3}{4}{\xi_{\sigma ss}^*}^2 + \frac{1}{4} {\xi_{\sigma \sigma \sigma}^*}^2 -3 \ep^* \eta_{ss}^* \right. \nonumber \\
&& \hspace{-2cm} \left. - {\ep^*}^2 +3 \ep^* \eta_{\sigma s}^* \cos 2 \theta^* + \frac{3}{2} (1-\sin 2\theta^*)(\eta_{ss}^* - \eta_{\sigma \sigma}^* + 2 \ep^*)\right) \,.
\eea
The term in square brackets is always negligible and this multiplies terms no larger than ${\cal O} ({\ep^*}^2)$ and so may be ignored, leaving the same simple result that we found for sum-separable potentials.

After these various terms have been neglected from Eqs. \eqref{eq:ss_tnl_full}
to \eqref{eq:ps_gnl_full}, we can simplify the remaining terms by rewriting them
by means of the trigonometric relations 
\bea
\tau_1 &=& \tau + 2 f + 1 \,, \nonumber \\
\tau_4 &=& 2 f_1 ( 1+ f)\,, \nonumber \\
\tau_7 &=& 4(\tau+f)\,, \nonumber \\
2 g_2 &=& \tau-f \,.
\eea

\bibliographystyle{JHEP}
\bibliography{gnl}

\end{document}